\documentclass[%
reprint,
superscriptaddress,
showpacs,preprintnumbers,
amsmath,amssymb,
aps,
prb,
floatfix,
notitlepage,
citeautoscript
]{revtex4-1}
\usepackage{array}
\usepackage{graphicx}
\usepackage[export]{adjustbox}
\usepackage{dcolumn}
\usepackage{bm}
\usepackage{amsmath} 
\usepackage{wasysym}
\usepackage[utf8]{inputenc}
\usepackage{color}
\usepackage{blindtext}
\usepackage{bm}
\usepackage{todonotes}
\usepackage{pgfplots}

\DeclareMathOperator{\sgn}{sgn}

\begin{document}
\newlength\fheight
\newlength\fwidth

\title{Monolayer transition metal dichalcogenides in strong magnetic fields: Validating the Wannier model using a microscopic calculation}

\author{J. Have}
\email{jh@nano.aau.dk}
\affiliation{Department of Materials and Production, Aalborg University, DK-9220 Aalborg East, Denmark}
\affiliation{Department of Mathematical Sciences, Aalborg University, DK-9220 Aalborg East, Denmark}
\author{G. Catarina}
\affiliation{QuantaLab, International Iberian Nanotechnology Laboratory (INL), 4715-330 Braga, Portugal}
\author{T.G. Pedersen}
\affiliation{Department of Materials and Production, Aalborg University, DK-9220 Aalborg East, Denmark}
\affiliation{Center for Nanostructured Graphene (CNG), DK-9220 Aalborg East, Denmark}
\author{N.M.R. Peres}
\affiliation{International Iberian Nanotechnology Laboratory (INL), 4715-330 Braga, Portugal}
\affiliation{Center and Department of Physics, and QuantaLab, University of Minho, Campus de Gualtar, 4710-057 Braga, Portugal}

\begin{abstract}
Using an equation of motion (EOM) approach, we calculate excitonic properties of monolayer transition metal dichalcogenides (TMDs) perturbed by an external magnetic field. We compare our findings to the widely used Wannier model for excitons in two-dimensional materials and to recent experimental results. We find good agreement between the calculated excitonic transition energies and the experimental results. In addition, we find that the exciton energies calculated using the EOM approach are slightly lower than the ones calculated using the Wannier model. Finally, we also show that the effect of the dielectric environment on the magnetoexciton transition energy is minimal due to counteracting changes in the exciton energy and the exchange self-energy correction.
\end{abstract}

\maketitle

\section{Introduction}
\label{sec:introduction}

The first use of an external magnetic field to study excitons and the electronic structure in thin film transition metal dichalcogenides (TMDs) was published in 1978\cite{tanaka1978excitons}. Since then, the study of magnetoexcitons has been an active field of research. With the recent emergence of monolayer TMDs, research in this area has undergone a rapid development, due in part to the interesting electronic and optical properties of monolayer TMDs\cite{xiao2012coupled, kormanyos2013monolayer, kormanyos2015k}, including large exciton binding energies on the order of 0.5-1 eV\cite{ramasubramaniam2012large, ramasubramaniam2012large, berkelbach2013theory, chaves2017excitonic}. Additionally, exciting magneto-optical phenomena of monolayer TMDs\cite{rose2013spin, chu2014valley, wang2017valley} have inspired novel applications, for which a detailed understanding of the effect of a magnetic field on the excitons is necessary. These phenomena include the valley Zeeman effect, a magnetic field assisted lifting of the degeneracy of the inequivalent $K$ and $K'$ valleys\cite{srivastava2015valley,aivazian2015magnetic,macneill2015breaking}. This control of the degeneracy could prove useful in the area of valleytronics\cite{schaibley2016valleytronics}. Another phenomenon lending itself to possible optical applications is Faraday rotation\cite{schmidt2011complex}, which has also been observed in monolayer TMDs perturbed by a magnetic field\cite{schmidt2016magnetic, da2018cavity}.

In addition to potential applications, perturbation by an external magnetic field provides experimental insight into the properties of excitons, such as their spatial extent\cite{stier2018magnetooptics,zipfel2018spatial} and the effect of the dielectric environment\cite{stier2016probing}. Using strong magnetic fields of up to 65 T, the Zeeman valley effect and diamagnetic shift of the excitonic states have been measured for the four most common monolayer TMDs: MoS\textsubscript{2}\cite{mitioglu2016magnetoexcitons,stier2016exciton}, MoSe\textsubscript{2}\cite{li2014valley,mitioglu2016magnetoexcitons,zipfel2018spatial}, WS\textsubscript{2}\cite{plechinger2016excitonic,stier2016magnetoreflection}, and WSe\textsubscript{2}\cite{stier2016probing,mitioglu2015optical}. The analysis of such experimental results would benefit from a thorough theoretical study of the effect of an external magnetic field on excitons. But while there is a plethora of experimental results on magnetoexcitons, there have been less theoretical studies. The difficulties related to a theoretical description of magnetoexcitons in two-dimensional materials is, in part, due to the magnetic field breaking the translation symmetry. In one-dimensional systems, translation symmetry can be retained by choosing a suitable gauge for the magnetic vector potential\cite{have2018magnetoexcitons}, but in two- and three-dimensional systems that option is not available. 

The standard theoretical approach has been to use an effective mass model such as the Wannier model\cite{wannier1937structure}, where the effective mass is calculated from the band structure of the unperturbed system. Using this approach, results regarding the binding energy of excitons, trions, and biexctions in monolayer TMDs perturbed by a magnetic field were recently published in Ref.~\onlinecite{van2018excitons}. But with no other theoretical models for magnetoexcitons in 2D materials, it can be difficult to validate the effective mass model. In addition, the effective mass model does not take into account the unique Landau level structure of monolayer TMDs\cite{rose2013spin, wang2017valley}, which affects the magneto-optical response. In this paper, we provide an alternative approach for describing magnetoexcitons, which does not depend on the effective mass approximation. The approach is an extension of the equation of motion (EOM) method in Ref.~\onlinecite{chaves2017excitonic} to the case, where the TMDs are perturbed by an external magnetic field. This model has several advantages, which include: Accounting for the Landau level structure of TMDs, allowing coupling between distinct bands and valleys, and providing a more self-contained theoretical framework. The EOM approach can also be used to calculate the optical response and was previously used to include second-order effects in the electric field in Ref.~\onlinecite{pedersen2015intraband}.

The present paper is structured as follows: In Sec.~\ref{sec:single-particle}, we introduce the single-particle Hamiltonian, which will serve as the outset for our study. In Sec.~\ref{sec:eom}, the EOM approach is briefly introduced. Sec.~\ref{sec:electron-electron} contains the definition of the electron-electron interaction Hamiltonian, as well as the derivation of the EOM for the excitonic problem. Sec.~\ref{sec:wannier} serves to introduce the Wannier model, which we will use for comparison with the results obtained in the EOM approach. Finally, in Sec.~\ref{sec:results} our results are presented and compared to recent experiments.

\section{Single-particle Hamiltonian}
\label{sec:single-particle}
\begin{figure}[t]
  \centering
  \includegraphics[width=0.49\textwidth]{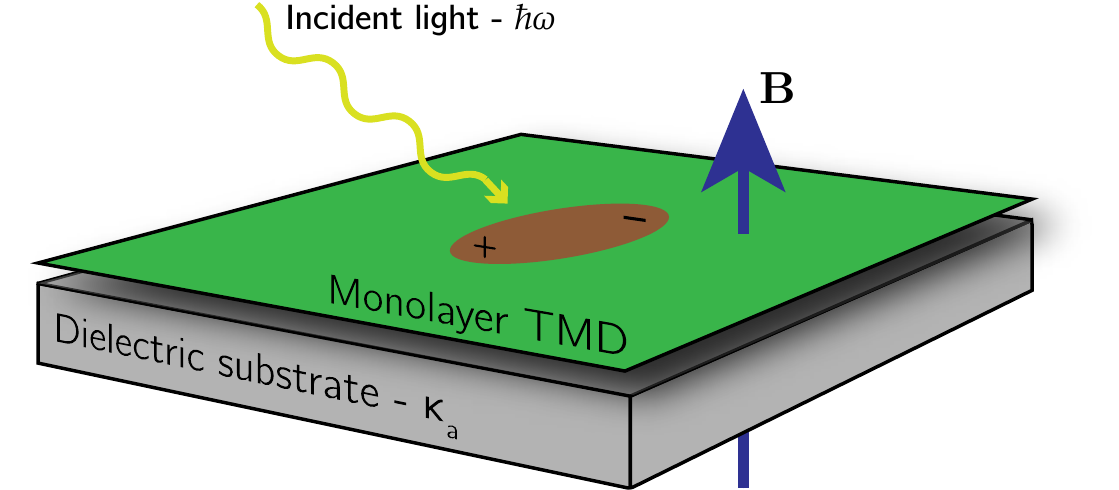}
  \caption{Sketch of the system under consideration: Excitons in a monolayer TMD material perturbed by a uniform static magnetic field perpendicular to the monolayer. The monolayer may be encapsulated between a dielectric substrate and a capping material.}\label{fig:geometry}
\end{figure}

In this section, we present the system and the single-particle Hamiltonian, which is the outset for our study of magnetoexcitons. The system is illustrated in Fig.~\ref{fig:geometry}. A monolayer TMD material, possibly deposited on some dielectric substrate with relative dielectric constant $\kappa_a$ and capped by a dielectric with relative dielectric constant $\kappa_b$, is perturbed by a uniform static magnetic field perpendicular to the TMD. Under absorption of an incident photon with energy $\hbar\omega$ an exciton is generated. The properties of the exciton, i.e. size and energy, are affected by the magnetic field. 

\begin{figure}[t]
  \centering
  \includegraphics[width=0.48\textwidth]{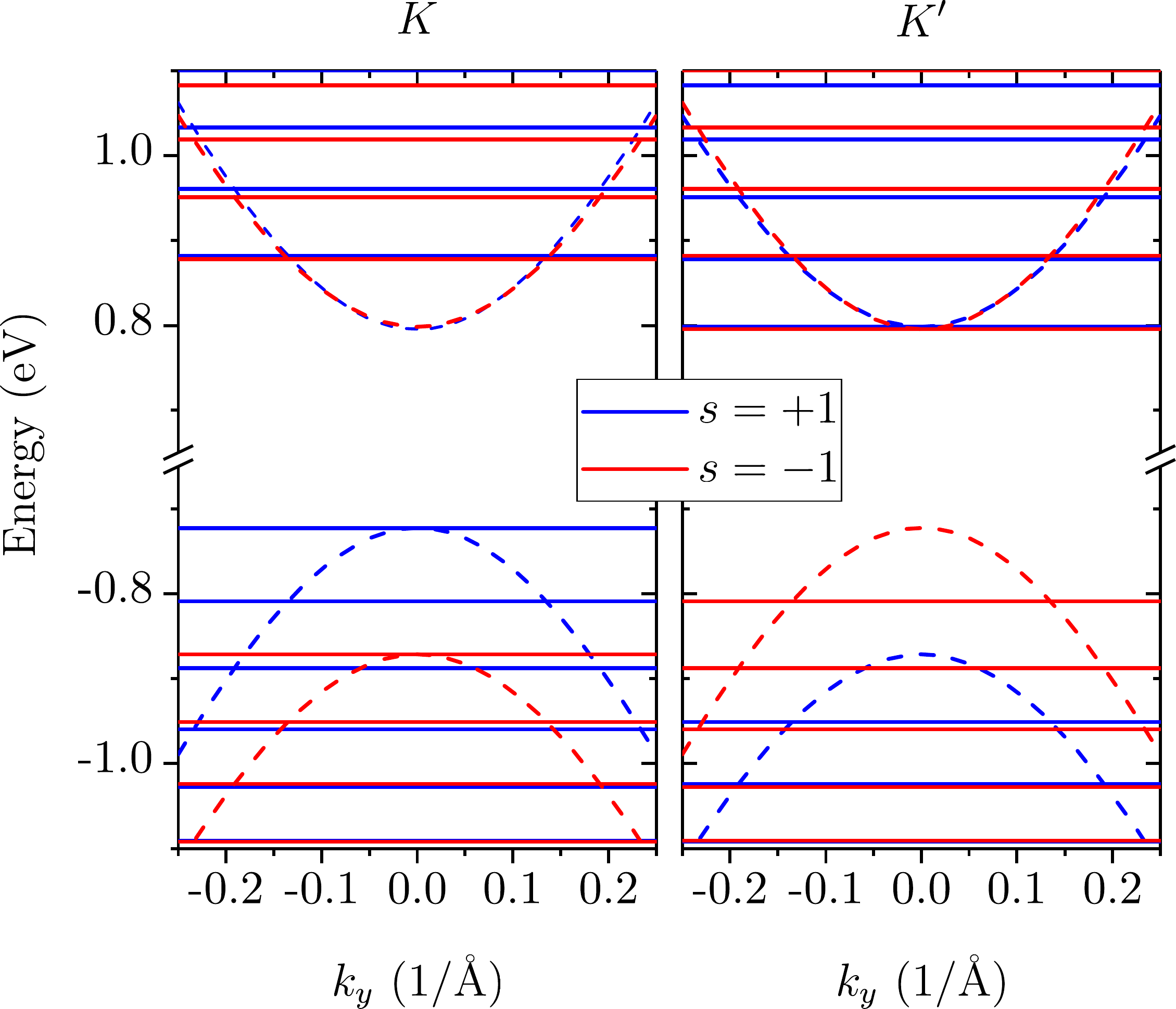}
  \caption{Single-particle spectrum at the $K$ and $K'$ valleys of MoS\textsubscript{2} with (solid lines) and without (dashed lines) magnetic field. Red and blue indicate spin up and spin down, respectively. The Landau level spectrum is plotted for a very high magnetic field (600 T) to make it possible to distinguish the individual Landau levels. Qualitatively similar features are found at lower magnetic field strengths.}\label{fig:energies}
\end{figure}

To describe magnetoexcitons in monolayer TMDs, we need an accurate description of the single-particle properties of unperturbed TMDs. For that purpose, we apply the effective Hamiltonian from Ref.~\onlinecite{kormanyos2013monolayer}. This effective Hamiltonian describes a massive Dirac system, and has been found to reproduce the band structure of monolayer TMDs in the low energy range around the direct band gaps in the $K$ and $K'$ valleys, including the spin-orbit splitting of the bands. For a monolayer in the $xy$-plane the Hamiltonian is given by
\begin{equation}\label{eq:H0}
  \hat{H}_0 = v_F(\tau \sigma_x p_x + \sigma_y p_y) + \Delta_{\tau, s}\sigma_z + \xi_{\tau, s}\mathbb{I},
\end{equation}
where $v_F$ is the Fermi velocity, $\tau=\pm 1$ is the valley index ($+1$ for the $K$ valley and $-1$ for the $K'$ valley), $\sigma_i$ are the Pauli matrices with $i\in\{x,y,z\}$, $p_x$ and $p_y$ are the canonical momentum operators, $\mathbb{I}$ is the $2\times 2$ identity matrix, and $\Delta_{\tau, s}$ and $\xi_{\tau, s}$ are the valley- and spin-dependent mass and onsite energy, respectively. The mass and onsite energy are given by
\begin{equation}
  \Delta_{\tau, s} = \Delta - \tau s \frac{\Lambda_1}{2}, \qquad \xi_{\tau, s} =  \tau s \frac{\Lambda_2}{2},
\end{equation}
where $s=\pm 1$ ($+1$ for the spin up and $-1$ for spin down), $\Lambda_1 = (\Delta_{\mbox{soc}}^{\mathcal{V}}-\Delta_{\mbox{soc}}^{\mathcal{C}})/2$ and $\Lambda_2 = (\Delta_{\mbox{soc}}^{\mathcal{V}}+\Delta_{\mbox{soc}}^{\mathcal{C}})/2$. The parameters $v_F$, $\Delta$, $\Delta_{\mbox{soc}}^{\mathcal{V}}$ and $\Delta_{\mbox{soc}}^{\mathcal{C}}$ are material dependent, and found by fitting to first principles band structure calculation\cite{xiao2012coupled,liu2013three}. The material parameters used in this paper are provided in Table~\ref{tab:parameters}. The single-particle energy bands are the eigenvalues $\varepsilon_{\tau, s}$ of $\hat{H}_0$, which are given by
\begin{equation}\label{eq:ss_eig}
    \varepsilon_{\tau, s} = \pm\sqrt{\hbar^2 v_F^2 |\mathbf{k}|^2 + \Delta_{\tau, s}^2} + \xi_{\tau, s}.
\end{equation}
Note that the eigenvalues only depend on the product $\tau s = \pm 1$, and not on $\tau$ and $s$ as individual parameters. The eigenvalues of MoS\textsubscript{2} are plotted as dashed lines in Fig.~\ref{fig:energies}. We observe that the energy dispersion shows spin-orbit splitting of both valence and conduction bands and that the $K$ and $K'$ valleys are inequivalent due to spin.

\begin{table}[b]
    \centering
    \begin{tabular}{l|c c c c }
     & $\Delta$ (eV) & $\hbar v_F$ (eV\AA\textsuperscript{-1}) & $\Delta_{\mbox{soc}}^{\mathcal{V}}$ (eV) & $\Delta_{\mbox{soc}}^{\mathcal{C}}$ (eV) \\ \hline\hline
    MoS\textsubscript{2} & 0.797 & 2.76 & 0.149 & -0.003 \\ \hline
    MoSe\textsubscript{2} & 0.648 & 2.53 & 0.186 & -0.022 \\ \hline
    WS\textsubscript{2} & 0.90 & 4.38 & 0.430 & 0.029 \\ \hline
    WSe\textsubscript{2} & 0.80 & 3.94 & 0.466 & 0.036 \\
  \end{tabular}\caption{Parameters of the effective Hamiltonian for the four common types of TMDs. The mass parameters and the Fermi velocities are taken from Ref.~\onlinecite{xiao2012coupled}, while the spin-orbit parameters are from Ref.~\onlinecite{liu2013three}.  An alternative set of parameters is provided in Ref.~\onlinecite{kormanyos2015k}.}
  \label{tab:parameters}
\end{table}

The next step is the inclusion of a perpendicular magnetic field $\mathbf{B}$. The magnetic field is introduced using the minimal coupling substitution $\mathbf{p} \mapsto \mathbf{p} + e\mathbf{A}$, where $\mathbf{p}$ is the momentum operator, $-e$ is the electron charge and $\mathbf{A}$ is the magnetic vector potential, related to the magnetic field by $\nabla \times \mathbf{A} = \mathbf{B}$. Using the Landau gauge, $\mathbf{A} = Bx\hat{\mathbf{y}}$, the effective perturbed Hamiltonian is
\begin{equation}
    \hat{H}_{B} = v_F\left[\tau \sigma_x p_x + \sigma_y(p_y + eBx)\right] + \Delta_{\tau,s}\sigma_z + \xi_{\tau,s}\mathbb{I}.
\end{equation}
The eigenvalues and eigenfunctions of $\hat{H}_B$ can be found by expressing $\hat{H}_B$ in terms of creation and annihilation operators\cite{rose2013spin,cheng2018nonlinear}, and then expanding the eigenfunctions in a basis of harmonic oscillator eigenfunctions. We find that the eigenvalues and the normalized eigenfunctions are given by
\begin{align}
  E^{n, \lambda}_{\tau, s} &= \lambda\sqrt{\Delta_{\tau, s}^2 + n(\hbar \omega_c)^2} + \xi_{\tau, s} \label{eq:ss-eigen},\\
  \Psi^{n, \lambda}_{\tau, s, k_y}(\mathbf{r}) &= \frac{e^{ik_y y}}{\sqrt{L_y}} \Phi^{n, \lambda}_{\tau, s}(\tilde{x})\label{eq:ss-wavefunction}.
\end{align}
Here, $n \geq (1+\tau \lambda)/2$ is the integer Landau level (LL) index, $\lambda = \pm$ indicates the type of LLs ($+$ for conduction type LLs and $-$ for valence type LLs),  $\hbar\omega_c = \sqrt{2}\hbar v_F / l_B$ is the cyclotron energy, $l_B = \sqrt{\hbar/(eB)}$ is the magnetic length, $L_y$ is the length of the system in the $y$ direction, and the spinor wavefunction is 
\begin{align}
  \Phi^{n, \lambda}_{\tau, s}(\tilde{x}) &= \frac{1}{\sqrt{2}}
  \left(\begin{array}{c}
  B_{\tau, s}^{n, \lambda}\phi_{n-(\tau + 1)/2}(\tilde{x}) \\
  C_{\tau, s}^{n,\lambda}\phi_{n+(\tau - 1)/2}(\tilde{x})
  \end{array}\right).
\end{align}
Here, $\tilde{x} = x + l_B^2 k_y$, $\phi_{n}(\tilde{x})$ are the usual harmonic oscillator eigenstates, and $B_{\tau, s}^{n, \lambda}$ and $C_{\tau, s}^{n,\lambda}$ are normalization constants given by
\begin{equation}
  B_{\tau, s}^{n, \lambda} = \lambda \sqrt{1+\lambda\alpha^{n}_{\tau, s}}, \quad C_{\tau, s}^{n, \lambda} = \sqrt{1-\lambda\alpha^{n}_{\tau, s}},
\end{equation}
where $\alpha^{n}_{\tau, s} = \Delta_{\tau, s}/\sqrt{\Delta_{\tau, s}^2 + n(\hbar \omega_c)^2}$. The harmonic oscillator eigenstates are given by
\begin{equation}
    \phi_n(\tilde{x}) = \frac{1}{\sqrt{2^n n!}}\left(\frac{1}{\pi l_B^2}\right)^{\frac{1}{4}}e^{-\frac{\tilde{x}^2}{2l_B^2}}H_n\left(\frac{\tilde{x}}{l_B}\right),
\end{equation}
where $H_n$ are the physicist's Hermite polynomials, which are defined by
\begin{equation}
    H_n(x) = (-1)^ne^{x^2}\frac{d^n}{dx^n}e^{-x^2}.
\end{equation}

Note that the energies $E^{n, \lambda}_{\tau, s}$ define a discrete set of LLs that have a degeneracy corresponding to the number of distinct $k_y$ values. The Landau level spectrum of MoS\textsubscript{2} is plotted (solid lines) in Fig. \ref{fig:energies}. From Fig.~\ref{fig:energies} and the allowed values of $n$, we see that a LL with $n=0$ is only allowed when $\tau \neq \lambda$. This gives rise to a magnetic-field-dependent increase of the band gap. Finally, the valley Zeeman splitting\cite{aivazian2015magnetic} is not included in the effective Hamiltonian $\hat{H}_B$. It could have been by adding additional terms to $\hat{H}_B$\cite{rybkovskiy2017atomically}, but since the focus of the present paper is on the excitonic effects, it is ignored for simplicity.  

\subsection{Dipole matrix elements}
\label{sec:dipole-elements}
In this section, the dipole matrix elements for the single-particle wavefunctions are calculated. In addition to being necessary for calculating the optical response, the dipole matrix elements provide information about the optical selection rules, which can be used to exclude some dark transitions from our excitonic calculations. This speeds up the numerical studies performed below by a significant factor. The interaction of the system with the incident light is included, within the dipole approximation, via the interaction Hamiltonian
\begin{equation}
    H_I = -\mathbf{d}\cdot \bm{\mathcal{E}}(t) = e \mathbf{r} \cdot \bm{\mathcal{E}}(t).
\end{equation}
Here, $\mathbf{d} = -e\mathbf{r}$ is the dipole moment operator and $\bm{\mathcal{E}}(t)$ the time-dependent electric field of the light. By construction, transitions between different valleys and different spins are not allowed. We introduce some notation to simplify the expressions. Let $\alpha$ be shorthand for $\{n, \lambda, k_y\}$ and $\eta$ for $\{\tau, s\}$, then the dipole matrix elements are written as  $\mathbf{d}^{\alpha \to \alpha'}_{\eta} = \langle \Psi^{n, \lambda}_{\tau, s, k_y}\left|\mathbf{d}\right|\Psi^{n', \lambda'}_{\tau' s', k_y'}\rangle$, where $\Psi^{n, \lambda}_{\tau, s, k_y}$ are the single-particle eigenstates of $\hat{H}_B$. For the dipole matrix elements in the $x$ direction, we find
\begin{align}
  d^{\alpha\to\alpha'}_{\eta, x} &= -e\delta_{k_y,k_y'}\left\langle \Phi^{n, \lambda}_{\tau, s}\left|x\right|\Phi^{n', \lambda'}_{\tau, s}\right\rangle \nonumber \\
&= -e\delta_{k_y,k_y'}\frac{\left\langle \Phi^{n, \lambda}_{\tau, s}\left|\left[\hat{H}_B,x\right]\right|\Phi^{n', \lambda'}_{\tau, s}\right\rangle}{E^{n, \lambda}_{\tau, s}-E^{n', \lambda'}_{\tau, s}}.
\end{align}
The commutator is simply $[\hat{H}_B,x] = -i\hbar v_F\tau\sigma_x$. A similar expression holds for the commutator with $y$. Consequently, the dipole matrix elements are found to be
\begin{align}
\mathbf{d}^{\alpha \to \alpha'}_{\eta} = \frac{e \hbar v_F \delta_{k_y,k_y'}}{2\Delta E^{n, \lambda}_{n', \lambda'}} & \left[ B_{\tau, s}^{\lambda, n}C_{\tau, s}^{n', \lambda'}\left(\begin{array}{c} -i \tau \\ 1\end{array}\right)\delta_{n-\tau,n'}\right. \nonumber\\
& - \left. B_{\tau, s}^{n', \lambda'}C_{\tau, s}^{n, \lambda}\left(\begin{array}{c} i \tau \\ 1\end{array}\right)\delta_{n+\tau,n'}\right].\label{eq:selection-rules}
\end{align}
Here, $\Delta E^{n, \lambda}_{n', \lambda'} := E^{n, \lambda}_{\tau, s}-E^{n', \lambda'}_{\tau, s}$. The non-zero dipole matrix elements correspond to the bright interband transitions. Equation~\eqref{eq:selection-rules} shows that the allowed interband transitions from a LL with index $n$ are to LLs with index $n'=n \pm 1$ and at the same $k_y$ points.

\section{Equation of motion approach}
\label{sec:eom}
The excitonic properties will be calculated using an EOM approach similar to that of Ref.~\onlinecite{chaves2017excitonic}, which is an extension of the method introduced to describe the magneto-optics of graphene in a cavity in Ref.~\onlinecite{ferreira2011faraday}. The approach relies primarily on writing and solving Heisenberg's equation of motion, which is given by
\begin{equation}\label{eq:Heisenberg}
  -i\hbar\frac{\partial \hat{\rho}}{\partial t} = [\hat{H}, \hat{\rho}].
\end{equation}
Here $\hat{H}$, is the full Hamiltonian including $\hat{H}_I$,  and $\hat{\rho}$ is the density matrix for the states of $\hat{H}_B$. 

To compute the density matrix, we introduce the creation and annihilation operators $\hat{c}^\dagger_{\alpha,\eta}(t)$ and $\hat{c}_{\alpha,\eta}(t)$, which, respectively, create or annihilate an electron in state $\Psi_{\alpha}^\eta \equiv \Psi_{\tau, s, k_y}^{n, \lambda}$ (Recall, that $\alpha$ is short for $\{n, \lambda, k_y \}$ and $\eta$ is short for $\{\tau, s\}$). The creation and annihilation operators obey the usual anti-commutator relations. Using these operators, we can express the single-particle Hamiltonian and the light-matter interaction Hamiltonian as
\begin{align}
    \hat{H}_B(t) &= \sum_{\alpha, \eta} E_{\alpha}^\eta \hat{\rho}_{\alpha,\alpha}^\eta(t), \\
    \hat{H}_I(t) &= -\bm{\mathcal{E}}(t) \cdot \sum_{\alpha, \alpha', \eta} \mathbf{d}_\eta^{\alpha \to \alpha'}\hat{\rho}_{\alpha,\alpha'}^\eta(t),
\end{align}
where $\hat{\rho}^\eta_{\alpha,\alpha'}(t) = \hat{c}^\dagger_{\alpha,\eta}(t)\hat{c}_{\alpha',\eta}(t)$ are elements of the density matrix in a basis of the eigenstates of $\hat{H}_B$.  Note that only a few of the terms in the sum over $\alpha'$ give non-zero contributions to $\hat{H}_I$ due to the optical selection rules from Sec.~\ref{sec:single-particle}.

Solving Heisenberg's EOM exactly as expressed in Eq.~\eqref{eq:Heisenberg} is not possible. Consequently, we take the expectation value on both sides of Eq.~\eqref{eq:Heisenberg} with respect to the equilibrium state, and get the following EOM for the expectation value
\begin{equation}\label{eq:eos-exp}
    -i\hbar\frac{\partial}{\partial t}p_{\alpha,\alpha'}^{\eta} = \left\langle[\hat{H}, \hat{\rho}_{\alpha,\alpha'}^\eta]\right\rangle,
\end{equation}
with $p_{\alpha,\alpha'}^\eta = \langle {\rho}_{\alpha,\alpha'}^\eta \rangle$. Note that the diagonal elements $\alpha=\alpha'$ define a new electron distribution. The commutators of $\hat{H}_B$ and $\hat{H}_I$ with the density matrix are calculated in Appendix~\ref{app:commutators} and can be used to calculate the single-particle optical response as in Ref.~\onlinecite{catarina2018magneto}. We now turn to the problem of including electron-electron interactions in the Hamiltonian and then find the excitonic states by solving Eq.~\eqref{eq:eos-exp}.

\section{Electron-electron interactions}
\label{sec:electron-electron}
From this point on, we consider the full Hamiltonian given by $\hat{H} = \hat{H}_B + \hat{H}_I + \hat{H}_{ee}$, where the electron-electron interaction Hamiltonian is defined by
\begin{equation}\label{eq:Hee}
\hat{H}_{ee} = \frac{1}{2}\int d\mathbf{r}_1 d\mathbf{r}_2 \hat{\psi}^\dagger(\mathbf{r}_1)\hat{\psi}^\dagger(\mathbf{r}_2)U(\mathbf{r}_1-\mathbf{r}_2)\hat{\psi}(\mathbf{r}_2)\hat{\psi}(\mathbf{r}_1).
\end{equation}
Here, the integrals also cover spin, $U(\mathbf{r})$ is the electron-electron interaction potential defined below, and $\hat{\psi}(\mathbf{r})$ is the field operator, given by
\begin{equation}\label{eq:field-operator}
    \hat{\psi}(\mathbf{r}) = \sum_{\substack{\alpha, \eta}}\hat{c}_{\alpha,\eta}\Psi_{\alpha}^\eta(\mathbf{r}).
\end{equation}
Here and in the following, we drop the explicit time dependence of $\hat{c}_{\alpha,\eta}(t)$ and $\hat{\rho}_{\alpha,\alpha'}^\eta(t)$ to simplify notation. 

In a strict two-dimensional system, the electron-electron interaction $U(\mathbf{r})$ is not the usual Coulomb potential, but instead given by the Keldysh potential\cite{keldysh1979coulomb}. In momentum space the Keldysh potential has the following simple form\cite{keldysh1979coulomb,cudazzo2011dielectric,trolle2017model}
\begin{align}\label{eq:keldysh}
  U(\mathbf{q}) = \frac{e^2}{2\varepsilon_0}\frac{1}{q(\kappa + r_0q)},
\end{align}
where $q=|\mathbf{q}|$, $\varepsilon_0$ is the vacuum permittivity, $r_0$ is a material dependent in-plane screening length, and $\kappa=(\kappa_a + \kappa_b)/2$ is the average of the relative dielectric constant of the substrate and the capping material. The in-plane screening lengths used in this paper are listed in Table~\ref{tab:mass_screening}.

Before calculating the commutator of $\hat{H}_{ee}$ with the density matrix and solving the Heisenberg EOM, we will rewrite $\hat{H}_{ee}$ slightly. Assuming that the electron-electron coupling between different valleys is negligible, the $\hat{H}_{ee}$ can be written as
\begin{align}\label{eq:electron-electron}
    \hat{H}_{ee} &= \frac{1}{2}\sum_{\substack{\tau, s, s' \\ \alpha_1, \alpha_2 \\ \alpha_3, \alpha_4}} U^{\tau, s, s'}_{\alpha_1 \alpha_ 4, \alpha_2 \alpha_3}\hat{c}^\dagger_{\alpha_1,\tau,s}\hat{c}^\dagger_{\alpha_2,\tau,s'}\hat{c}_{\alpha_3,\tau,s'}\hat{c}_{\alpha_4,\tau,s},
\end{align}
where two of the summations over spin cancel because of the spin integrals in Eq.~\eqref{eq:Hee}, and the so-called Coulomb integrals are
\begin{equation}\label{eq:coulomb-integrals}
    U^{\tau, s, s'}_{\alpha_1 \alpha_4, \alpha_2 \alpha_3} = \frac{1}{4\pi^2}\int\mathrm{d}^2\mathbf{q}~U(\mathbf{q})F_{\alpha_1,\alpha_4}^{\tau, s}(\mathbf{q})F_{\alpha_2,\alpha_3}^{\tau, s'}(-\mathbf{q}).
\end{equation}
Here, $F_{\alpha,\alpha'}^{\tau, s}(\mathbf{q})$ are structure factors defined as
\begin{equation}\label{eq:structure-factors}
    F_{\alpha,\alpha'}^{\tau, s}(\mathbf{q}) = \int\mathrm{d}^2\mathbf{r}~e^{i\mathbf{q}\cdot \mathbf{r}} (\Psi^{\alpha}_{\tau, s}(\mathbf{r}))^{*}\Psi^{\alpha'}_{\tau, s}(\mathbf{r}).
\end{equation}
An explicit expression for the structure factors is provided in Appendix~\ref{app:formfactors}. Using Eq.~\eqref{eq:electron-electron}, we calculate the commutator of the full Hamiltonian with the density matrix in Appendix~\ref{app:commutators} and find that the EOM in Eq.~\eqref{eq:eos-exp} can be written as
\begin{widetext}
    \begin{align}\label{eq:complete-eom}
        \left(E_{\alpha'}^\eta-E_{\alpha}^\eta-i\hbar\frac{\partial }{\partial t}\right) p^\eta_{\alpha, \alpha'} =   \sum_{\substack{\alpha_1,\alpha_2 \\ \alpha_3}} p_{\alpha_1, \alpha_3}^\eta \left( U^{\tau, s, s}_{\alpha' \alpha_3, \alpha_1 \alpha_2}p_{\alpha,\alpha_2}^\eta - U^{\tau, s, s}_{\alpha_1 \alpha, \alpha_2 \alpha_3}p_{\alpha_2,\alpha'}^\eta\right)-\bm{\mathcal{E}}(t)\cdot\sum_{\alpha''} \left(  \mathbf{d}_\eta^{\alpha'' \to \alpha}p_{\alpha'',\alpha'}^\eta - \mathbf{d}_\eta^{\alpha' \to \alpha''}p_{\alpha,\alpha''}^\eta\right).
    \end{align}
\end{widetext}    
Here, $E_{\alpha}^\eta \equiv E_{n,\lambda}^{\tau, s}$ and the expectation value of the four-body operator in $\hat{H}_{ee}$ has been truncated at the random phase approximation (RPA) level\cite{ehrenreich1959self}. Comparing the EOM to what was found in Ref.~\onlinecite{chaves2017excitonic}, we see that the general form of the equation is equivalent to the expression for a system with an arbitrary number of bands. In the following subsections, we keep only the terms of Eq.~\eqref{eq:complete-eom}, which are of first order in the electric field and collect the terms corresponding to the exchange self-energy corrections and electron-hole interactions.

\subsection{Exchange self-energy corrections}
\label{sec:exchange-self-energy}
In this section, we briefly touch upon the exchange self-energy corrections caused by the electron-electron interactions. The term exchange should be understood in the sense of the Hartree-Fock approximation, where there are two corrections to self-energy: The Hartree correction, which is canceled by the interaction with the positive background (see Appendix~\ref{app:commutators}), and the exchange correction. 

Although exchange self-energy corrections are not the main focus of this work, it is still important to include them if we hope to accurately describe the transition energy of the excitons. This is because the self-energy correction has a strong impact on the value of the single particle gap. In Appendix~\ref{app:commutators}, the first order terms that result in a renormalization of the LLs are collected. It is found that the self-energy renormalized LLs, $\tilde{E}^\eta_{\alpha}$, are given by
\begin{equation}
    \tilde{E}^{\eta}_\alpha = E^{\eta}_\alpha - \Sigma_{\alpha}^\eta, \qquad \Sigma_\alpha^\eta = \sum_{\alpha'}f(E_{\alpha'}^\eta) U_{\alpha'\alpha, \alpha \alpha'}^{\tau, s, s}.
\end{equation}
Here, $\Sigma_{\alpha}^\eta$ is the exchange self-energy correction and $f(E)$ is the Fermi-Dirac distribution. We calculate the exchange self-energy correction using the structure factors from Appendix~\ref{app:formfactors}. Converting the sum over $k_y$ to an integral, the exchange self-energy can be written as
\begin{equation}\label{eq:xc_self_energy}
    \sum_{\alpha'}U^{\tau, s, s}_{\alpha' \alpha, \alpha \alpha'}f(E_{\alpha'}^{\eta}) = \sum_{n',\lambda'}f(E_{n',\lambda'}^{\eta})I_{\lambda'n',\lambda n}^{\eta},
\end{equation}
where the integrals are defined as
\begin{equation}\label{eq:xc-integrals}
  I_{\lambda n,\lambda' n'}^\eta = \frac{1}{16\pi^2}\int \mathrm{d}^2\mathbf{q} ~U\left(\mathbf{q}\right)e^{-\frac{l_B^2q^2}{2}}\left|J_{\lambda n,\lambda' n'}^{\eta}(\mathbf{q})\right|^2.
\end{equation}
Here, $J_{\lambda n,\lambda' n'}^{\eta}$ is the function defined in Eq.~\eqref{eq:J_function}. The integral in Eq.~\eqref{eq:xc-integrals} is simplified by the fact that $U(\mathbf{q})$ and $|J_{\lambda n,\lambda' n'}^{\eta}(\mathbf{q})|^2$ only depend on $q=|\mathbf{q}|$, meaning that the angular integral simply gives a factor of $2\pi$. In the remainder of the paper, we assume that the system is undoped, i.e. the Fermi level is in the band gap, and that $T = 0$ K. This implies that the sum in Eq.~\eqref{eq:xc_self_energy} only runs over the valence type LLs, which simplifies the numerical calculations. 

For graphene described in the Dirac approximation, the exchange self-energy correction has been found to diverge logarithmically when summing over an infinite number of valence LLs\cite{shizuya2010many}. We have observed the same type of divergence numerically for the expression in Eq.~\eqref{eq:xc_self_energy}. Consequently, a cutoff $N_{cut}$ of the summation over LLs has to be introduced. In Ref.~\onlinecite{sokolik2017many}, (see also Ref.~\onlinecite{nilsson2006electronic}) the cutoff was calculated for graphene by equating the concentration of electrons in $N_{cut}$ LLs to that in the filled valence band. The same approach can be used for TMDs and we find a cutoff equal to
\begin{equation}\label{eq:cutoff}
    N_{cut} = \frac{\pi l_B^2}{\Omega_0},
\end{equation} 
with $\Omega_0 = \sqrt{3}a^2/2$ the area of the primitive unit cell of the TMD. Taking $a=3.2$ \AA ~for all four TMDs\cite{rasmussen2015computational}, we get a cutoff equal to $N_{cut} \approx 2.33 \times 10^4 / B$ T.

\subsection{Excitonic effects}
Finally, using the exchange self-energy corrected LLs, we proceed to calculating the excitonic effects of TMDs perturbed by an external magnetic field. As shown in Appendix~\ref{app:commutators}, the excitonic states can be found by solving the first-order equation
\begin{align}\label{eq:excitonic-equation}
    &\left(\tilde{E}^\eta_{\alpha'}-\tilde{E}^\eta_\alpha -i\hbar\frac{\partial }{\partial t}\right) p^{\eta,1}_{\alpha, \alpha'} = \nonumber \\ 
    & \quad \left(\sum_{\alpha_1,\alpha_2}U_{\alpha' \alpha_2, \alpha_1 \alpha}^{\tau, s,s}p_{\alpha_1,\alpha_2}^{\eta, 1} - \bm{\mathcal{E}}(t)\cdot \mathbf{d}_{\eta}^{\alpha' \to \alpha}\right)\Delta f_{\alpha',\alpha}^{\eta},
\end{align}
where $\Delta f_{\alpha',\alpha}^{\eta} = f(E_{\alpha'}^\eta)-f(E_\alpha^\eta)$. As in Ref.~\onlinecite{chaves2017excitonic} the excitonic transition energies can be calculated by solving the homogeneous equation, i.e. setting $\bm{\mathcal{E}}(t) = \mathbf{0}$. Changing from time to frequency domain, we get the homogeneous equation
\begin{equation}\label{eq:homogeneous}
    (\tilde{E}^{\eta}_{\alpha'} - \tilde{E}^{\eta}_{\alpha} - E) p^{\eta}_{\alpha,\alpha'} =  \sum_{\alpha_1, \alpha_2}U^{\tau, s, s}_{\alpha'\alpha_2, \alpha_1\alpha}p^{\eta}_{\alpha_1, \alpha_2}  \Delta f_{\alpha',\alpha}^{\eta}.
\end{equation}
Here, $p^{\eta}_{\alpha,\alpha'}$ should be understood as the Fourier transform of $p^{\eta, 1}_{\alpha,\alpha'}$ and $E$ is the exciton transition energy for a fixed combination of spin and valley. The excitonic states are the interband solutions of Eq.~\eqref{eq:homogeneous}, i.e. where $\alpha$ and $\alpha'$ correspond to valence and conduction states, respectively. Thus, we assume that to be the case. Additionally, the sum over $\alpha_1$ and $\alpha_2$ can be split into two contributions: One where $\alpha_1$ and $\alpha_2$ are valence and conduction states, respectively, and one where the converse holds. We denote these cases the resonant contribution and the non-resonant contribution, respectively. In the following, we keep only the resonant contribution. It has been shown in Ref.~\onlinecite{chaves2017excitonic} that this is a valid approximation.

\begin{figure}[t]
    \centering
    \includegraphics[width=0.48\textwidth]{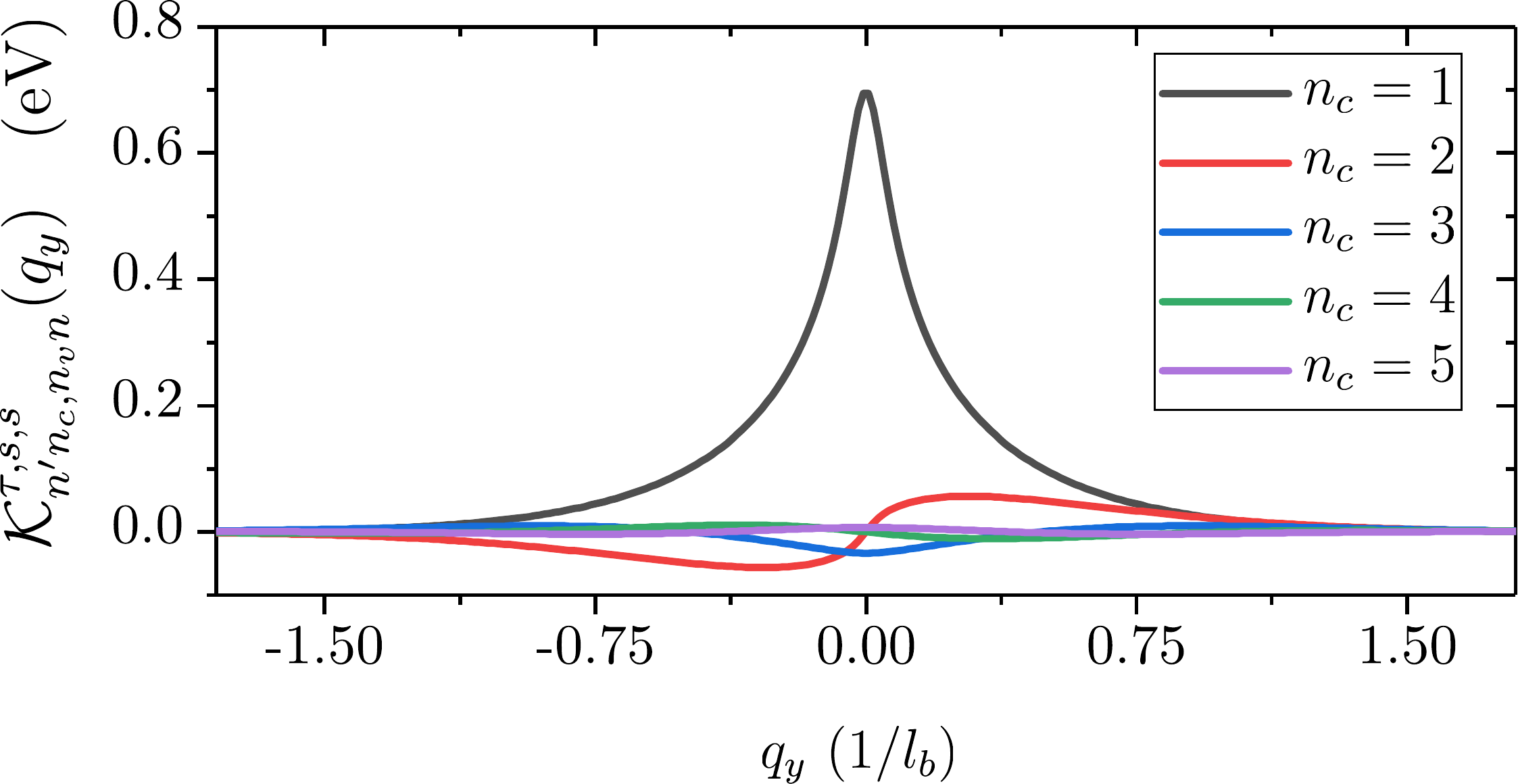}
    \caption{Electron-hole interaction kernels plotted for MoS\textsubscript{2} in a magnetic field of 100 T. The kernels are plotted for the $K$ valley, spin up, and $(n, n', n_v)=(0, 1, 0)$.}\label{fig:kernel}
\end{figure}

To clearly distinguish the valence and conduction states, we write $\alpha_v$ and $\alpha_c$ for $\alpha_1$ and $\alpha_2$ in Eq.~\eqref{eq:homogeneous}, respectively. Setting $k_y=k_y'$ (which corresponds to ignoring the dark non-vertical transitions, see Sec.~\ref{sec:dipole-elements}), we simplify the right hand side of Eq.~\eqref{eq:homogeneous} by writing
\begin{align}\label{eq:almost-kernel}
    \sum_{\alpha_v,\alpha_c}U^{\tau,s,s}_{\alpha' \alpha_c, \alpha_v \alpha}p^{\eta}_{\alpha_v,\alpha_c} \approx \sum_{n_v, n_c}\int_{-\infty}^\infty& \mathrm{d}q_y ~ \mathcal{K}_{n' n_c, n_v n}^{\tau, s, s}(q_y-k_y) \nonumber \\
    &\times p^{\eta}_{n_v, n_c}(q_y).
\end{align}
Here, we write the approximate sign to indicate the approximations discussed above, and, we denote $p_{\alpha_v,\alpha_c}^\eta$ by $p^\eta_{n_v,n_c}(k_y)$ in the case, where the $k_y$ values associated with $\alpha_c$ and $\alpha_v$ are equal. The different $\lambda$ parameters are fixed by the previous assumptions and are not written explicitly. The electron-hole interaction kernel $\mathcal{K}^{\tau, s, s}_{n' n_c, n_v n}$ is calculated using the structure factors and is found to be
\begin{align}
    \mathcal{K}_{n' n_c, n_v n}^{\tau, s, s}(q_y) = &\frac{1}{16\pi^2}\int_{-\infty}^{\infty} \mathrm{d}q_x ~ U\left(\mathbf{q}\right)e^{-\frac{l_B^2|\mathbf{q}|^2}{2}} \nonumber \\ &\times J^{\tau, s}_{+n', +n_c}(\mathbf{q})J^{\tau, s}_{-n_v, -n}(-\mathbf{q}),
\end{align}
where the integral over $q_x$ must be performed numerically. This finally implies a homogeneous first-order equation given by
\begin{align}\label{eq:BSE}
    &(\tilde{E}^\eta_{\alpha'}-\tilde{E}^\eta_\alpha-E)p^\eta_{n, n'}(k_y) \nonumber\\ 
    & \quad =\sum_{n_v, n_c}\int_{-\infty}^\infty\mathrm{d}q_y~\mathcal{K}_{n' n_c, n_v n}^{\tau, s, s}(q_y-k_y)p_{n_v,n_c}^\eta(q_y).
\end{align}
Equation~\eqref{eq:BSE} corresponds to the Bethe-Salpeter equation for electron-hole pairs\cite{rohlfing2000electron}, and it can be written as an eigenvalue problem with eigenvalues $E$ by discretizing the integral over $q_y$. The size of the eigenvalue problem scales as $N_k N_c N_v$, where $N_k$ is the number of points used to discretize the integral, and where $N_c$ and $N_v$ are the number of conduction and valence LLs, respectively. It is clear that only if the electron-hole kernel decays sufficiently fast with increasing $n_c$ and $n_v$ can we hope to solve Eq.~\eqref{eq:BSE}, since that would imply that the sums over $n_c$ and $n_v$ can be truncated. Fortunately, the kernel does decay quite fast in $n_c$ and $n_v$, as illustrated for $n_c$ in Fig.~\ref{fig:kernel}. In the next section, we turn our attention to an alternative (and non-microscopic) description of the excitonic properties of TMDs.

\begin{table}[t]
    \centering
    \begin{tabular}{l|c c c }
     & $\mu_{\tau, +1}$ ($m_e$) & $\mu_{\tau, -1}$ ($m_e$) & $r_0$ (\AA) \\ \hline\hline
    MoS\textsubscript{2} & 0.380 & 0.418 & 41.4 \\ \hline
    MoSe\textsubscript{2} & 0.355 & 0.417 & 51.7 \\ \hline
    WS\textsubscript{2} & 0.159 & 0.199 & 37.9  \\ \hline
    WSe\textsubscript{2} & 0.170 & 0.223 & 45.1 \\
  \end{tabular}\caption{Parameters used in the calculation of the excitonic properties for the four common types of TMDs. The first and second column contain the reduced exciton masses for the spin up and down bands, respectively. The third column is the in-plane screening length, and is taken from Ref.~\onlinecite{berkelbach2013theory}.}
  \label{tab:mass_screening}
\end{table}

\section{Wannier Model}
\label{sec:wannier}
In this section, we briefly introduce the Wannier model\cite{wannier1937structure} for excitons. The Wannier model is based on the effective mass approximation for a single pair of valence and conduction bands. For a two-dimensional semiconductor in a perpendicular magnetic field (using the symmetric gauge for the magnetic vector potential), the operator describing zero angular momentum excitons, i.e. $s$-type states, is\cite{stafford1990nonlinear}
\begin{equation}\label{eq:H_wannier}
  \hat{H}_{ex} = -\frac{\hbar^2}{2\mu}\nabla^2 + \frac{e^2B^2}{8\mu}r^2 - U(\mathbf{r}).
\end{equation}
Here, $\mu$ is the reduced effective mass, $\nabla^2$ is the 2D Laplace operator, $r$ is the relative electron-hole distance, and $U(\mathbf{r})$ is the electron-hole interaction potential given as the real space representation of Eq.~\eqref{eq:keldysh}. Taking the inverse Fourier transform of Eq.~\eqref{eq:keldysh}, we find
\begin{equation}\label{eq:keldysh_real}
  U(\mathbf{r}) = \frac{e^2}{8\varepsilon_0 r_0}\left[H_0\left(\frac{\kappa r}{r_0}\right)-Y_0\left(\frac{\kappa r}{r_0}\right)\right],
\end{equation}
with $r=|\mathbf{r}|$, $H_0$ the Struve function and $Y_0$ a Bessel function of the second kind.

\begin{figure*}[t]
    \centering
    \includegraphics[width=0.95\textwidth]{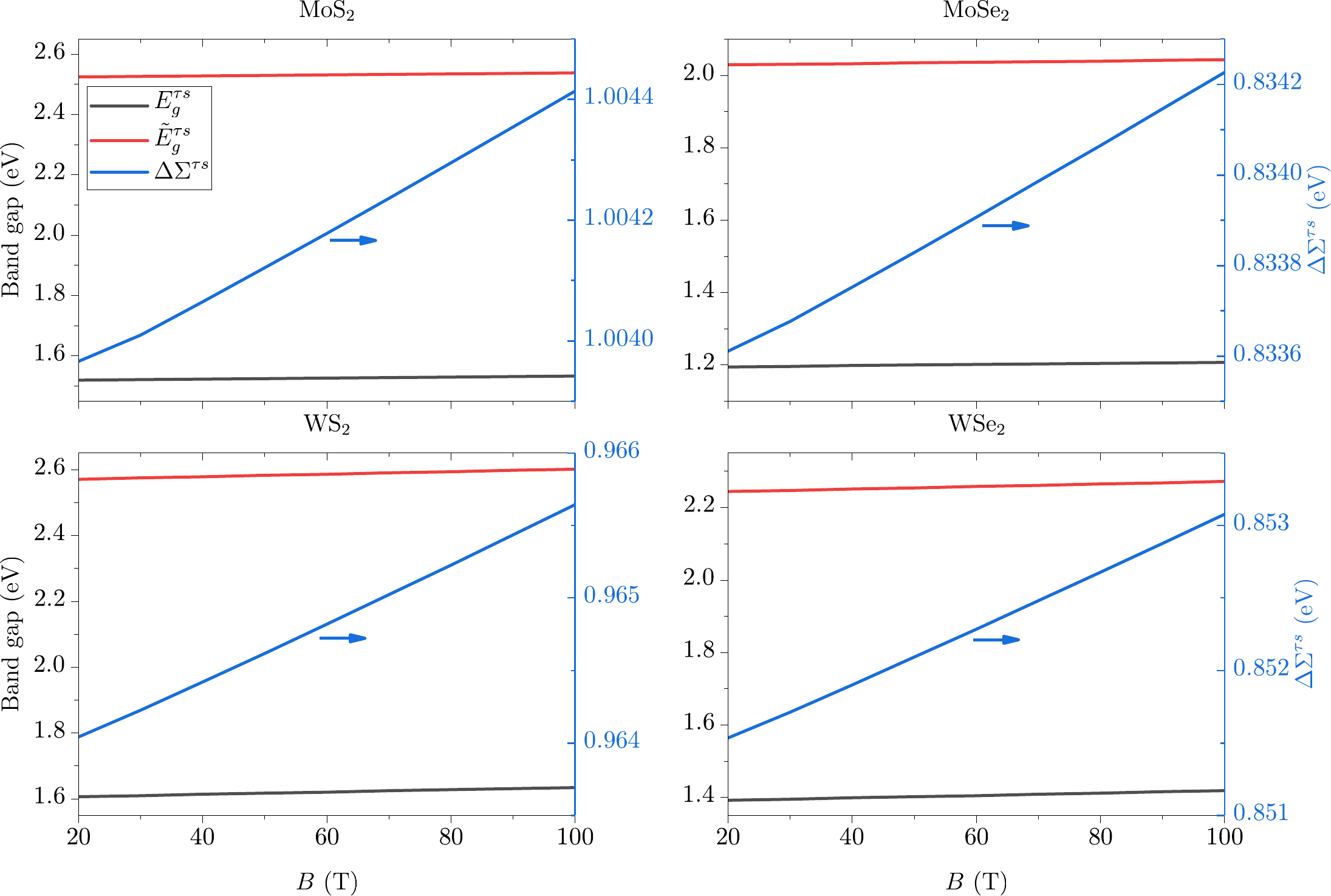}
    \caption{Plot of $\tau s= +1$ band gaps of suspended monolayer TMDs, i.e. taking $\kappa=1$. The uncorrected (black) and exchange self-energy corrected (red) band gaps are shown as a function of magnetic field. In addition, the exchange self-energy correction to the band gaps, $\Delta\Sigma^{\tau s} \equiv \tilde{E}^{\tau s}_g - E^{\tau s}_g$, is plotted (blue). The blue lines refer to the blue axes, while the rest refer to the black axes.}\label{fig:bandgaps}
\end{figure*}

For a direct comparison of the Wannier model with the solutions to Eq.~\eqref{eq:BSE}, we want to use the same parameters in both models. Thus, we calculate the effective mass from the eigenvalues of the unperturbed single-particle operator $\hat{H}_0$. Expanding the eigenvalues in Eq.~\eqref{eq:ss_eig} around $|\mathbf{k}|=0$, we find that the effective mass of an electron or hole in the $\tau$ valley and with spin $s$ is
\begin{equation}
    m^{*}_{\tau, s} = \frac{|\Delta_{\tau, s}|}{v_F^2}.
\end{equation}
The effective masses of electrons and holes are equal due to the symmetric conduction and valence bands. The reduced effective mass is then $\mu_{\tau, s} = m^{*}_{\tau, s}/2$, for which the values for the four common TMDs are given in Table~\ref{tab:mass_screening}. 

The $s$-type excitons, corresponding to bright excitons\cite{pedersen2016exciton}, can be found by solving the eigenvalue problem $\hat{H}_{ex} \psi(r) = E_{exc}\psi(r)$, where $E_{exc}$ is the exciton energy. We solve it by expanding $\psi(r)$ in a basis of Bessel functions, more specifically the basis $\phi_{i}(r) = J_0(\lambda_{i}r/R)$, where $\lambda_{i}$ is the $i$'th zero of the Bessel $J_0$ function and $r\leq R$. This basis corresponds to introducing an infinite barrier at $r=R$, but this should not affect the results as long as $R$ is sufficiently large. The same basis was recently used to describe the Stark shift of excitons in monolayer TMDs\cite{pedersen2016exciton,massicotte2018dissociation}.

\section{Results}
\label{sec:results}
In this section, our results are presented and discussed. In addition, we devote some attention to the computational approaches applied. All results were obtained using the parameters in Tabs.~\ref{tab:parameters} and \ref{tab:mass_screening}. Evaluating the integrals in the exchange self-energy correction, i.e. Eq.~\eqref{eq:xc-integrals}, is done using an adaptive quadrature and a numerical high precision library\cite{mpmath}. This approach, although computationally expensive, is found to provide accurate results for the rapidly oscillating integrands that occur when $n$ and $n'$ are large. In contrast, since the sum in Eq.~\eqref{eq:BSE} can be truncated at reasonably low values of $n_c$ and $n_v$, as illustrated by Fig.~\ref{fig:kernel}, the integral in the electron-hole kernel can be evaluated using the Gauss-Hermite quadrature. For the calculation of excitonic energies using the Wannier model, we use 400 basis functions and fix $R$ at $R = 20$ nm. The kinetic and magnetic matrix elements can be calculated analytically in this basis, while the potential matrix elements are computed numerically using a Gauss-Legendre quadrature.

First, we consider the exchange corrections. We denote the exchange self-energy corrected and the uncorrected band gaps as $\tilde{E}_g^{\tau s}$ and $E_g^{\tau s}$, respectively. The $\tilde{E}_g^{\tau s}$ and $E_g^{\tau s}$ band gaps are plotted in Fig.~\ref{fig:bandgaps} as a function of magnetic field for $\tau s = +1$ , i.e. spin up at the $K$ valley or spin down at the $K'$ valley. The results show that the self-energy correction gives rise to an opening of the band gap on the order of 0.8 to 1.0 eV. Similar values hold for the $\tau s = -1$ gaps. We find smaller exchange self-energy corrections than those of Ref.~\onlinecite{chaves2017excitonic} for the case of unperturbed monolayer TMDs. The explanation for this discrepancy is twofold: Firstly, we use a different parameter set. Secondly, the cutoffs that are used are different. But, as will be shown later, our approach results in exciton transition energies that match experiments quite well. 

\begin{figure}[t]
    \centering
    \includegraphics[width=0.48\textwidth]{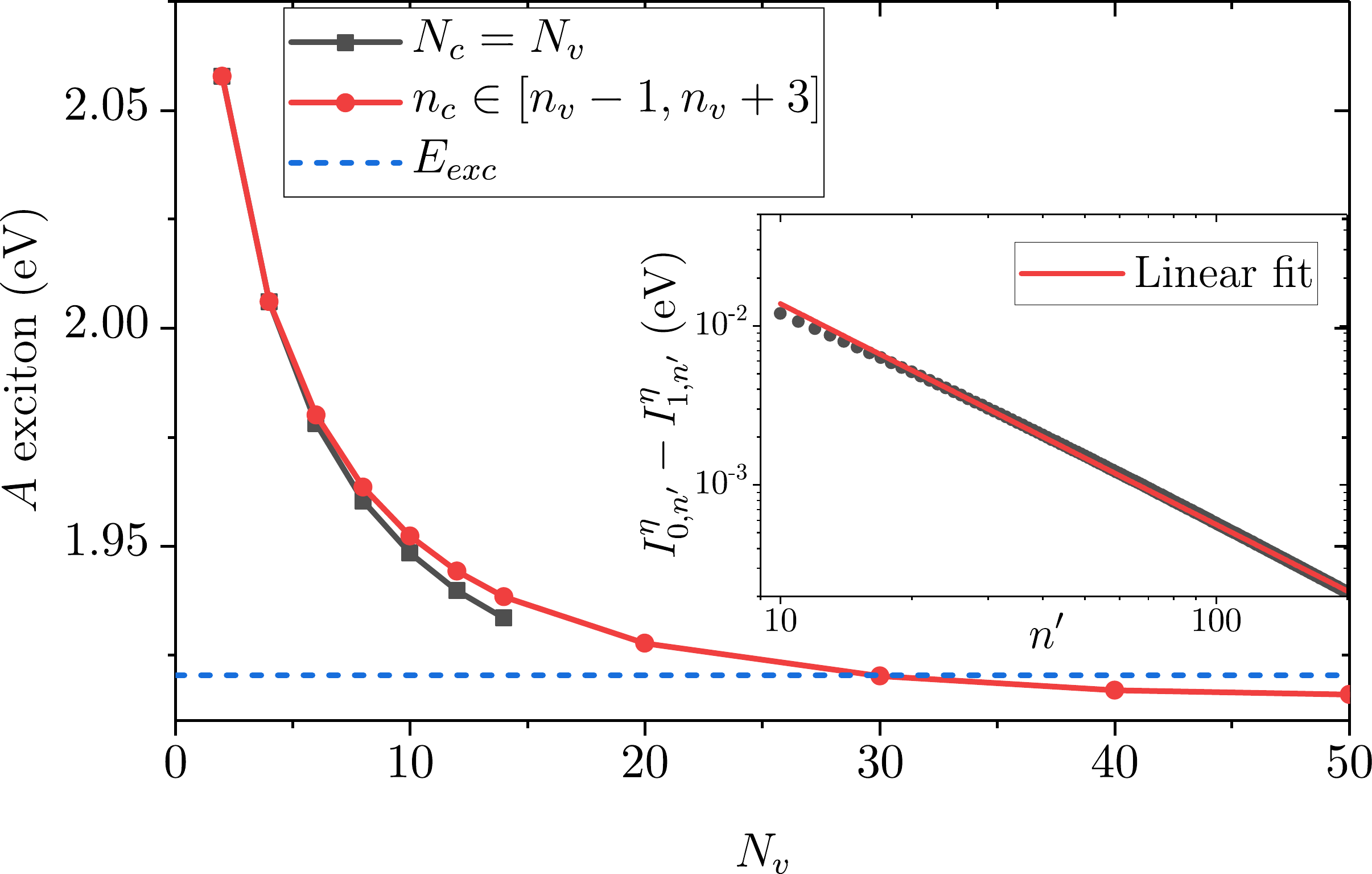}
    \caption{Convergence of the transition energy of the $A$ exciton in MoS\textsubscript{2} in a 100 T field. The black line refers to the situation where all LLs up to a cutoff $N_v=N_c$ are included and the red line refers to the situation where only significant transitions are included, i.e. of the type $n_v$ to $n_c \in [n_v - 1,n_v+3]$. The dashed blue line is the exciton transition energy calculated. Finally, the inset shows the integrals $I^{\eta}_{-0,-n'}-I^\eta_{+1,-n'}$ on a $\log$-$\log$ scale for $\tau s = +1$.}\label{fig:convergence}
\end{figure}

Considering the magnetic field dependence of the band gaps, we see that the uncorrected band gaps calculated using the LL energies in Eq.~\eqref{eq:ss-eigen} vary linearly with magnetic field for the field range in Fig.~\ref{fig:bandgaps}. We also find a linear magnetic field dependence of the exchange self-energy correction to the band gap with slopes of $ 5.57 ~\mu\mbox{eV}/\mbox{T}$ for MoS\textsubscript{2}, $7.76~ \mu\mbox{eV}/\mbox{T}$ for MoSe\textsubscript{2}, $20.0~ \mu\mbox{eV}/\mbox{T}$ for WS\textsubscript{2}, and $19.3~ \mu\mbox{eV}/\mbox{T}$ for WSe\textsubscript{2}. The slopes are for $\tau s = +1$ states, but similar slopes hold for the $\tau s = -1$ states. This apparent linear behavior of $\Delta\Sigma^{\tau s} = \tilde{E}_g^{\tau s}-E_g^{\tau s}$ can be explained by studying the expression in Eq.~\eqref{eq:xc_self_energy}. For small $B$, it can be shown using Eqs.~\eqref{eq:xc-integrals} and \eqref{eq:J_function} that the integrals $I^{\eta}_{\lambda n, \lambda' n'}$ are proportional to $\sqrt{B}$, for all $\lambda, \lambda',n$ and $n'$. If we can show that $I^{\eta}_{-0,-n'}-I^\eta_{+1,-n'}$ is proportional to $(n'+1)^{-3/2}$ as a function of $n'$, the result is a linear behavior of $\Delta\Sigma^{\tau s}$ since
\begin{align}
    \Delta\Sigma^{\tau s} &\propto \sqrt{B}\sum_{n'}^{N_{cut}}(n'+1)^{-\frac{3}{2}} \approx \sqrt{B}\int_1^{N_{cut}+1}\mathrm{d}n'~{n'}^{-\frac{3}{2}} \nonumber \\
    & \approx 2\sqrt{K}B.
\end{align}
Here, the last approximation holds for a cutoff of the type $N_{cut} = K/B$, with $K$ some constant, and for small $B$. The inset in Fig.~\ref{fig:convergence} shows $I^\eta_{-0,-n'}-I^\eta_{+1,-n'}$ on a $\log$-$\log$ scale for MoS\textsubscript{2}, with $B=100$ T and $\tau s = +1$. Fitting with a linear function, we find a power of $q=-1.33 \pm 0.03$ covering the range from 20 $T$ to 100 $T$. Thus, an approximately linear behavior of the exchange self-energy correction is expected.

\begin{figure}[b]
    \centering
    \includegraphics[width=0.49\textwidth]{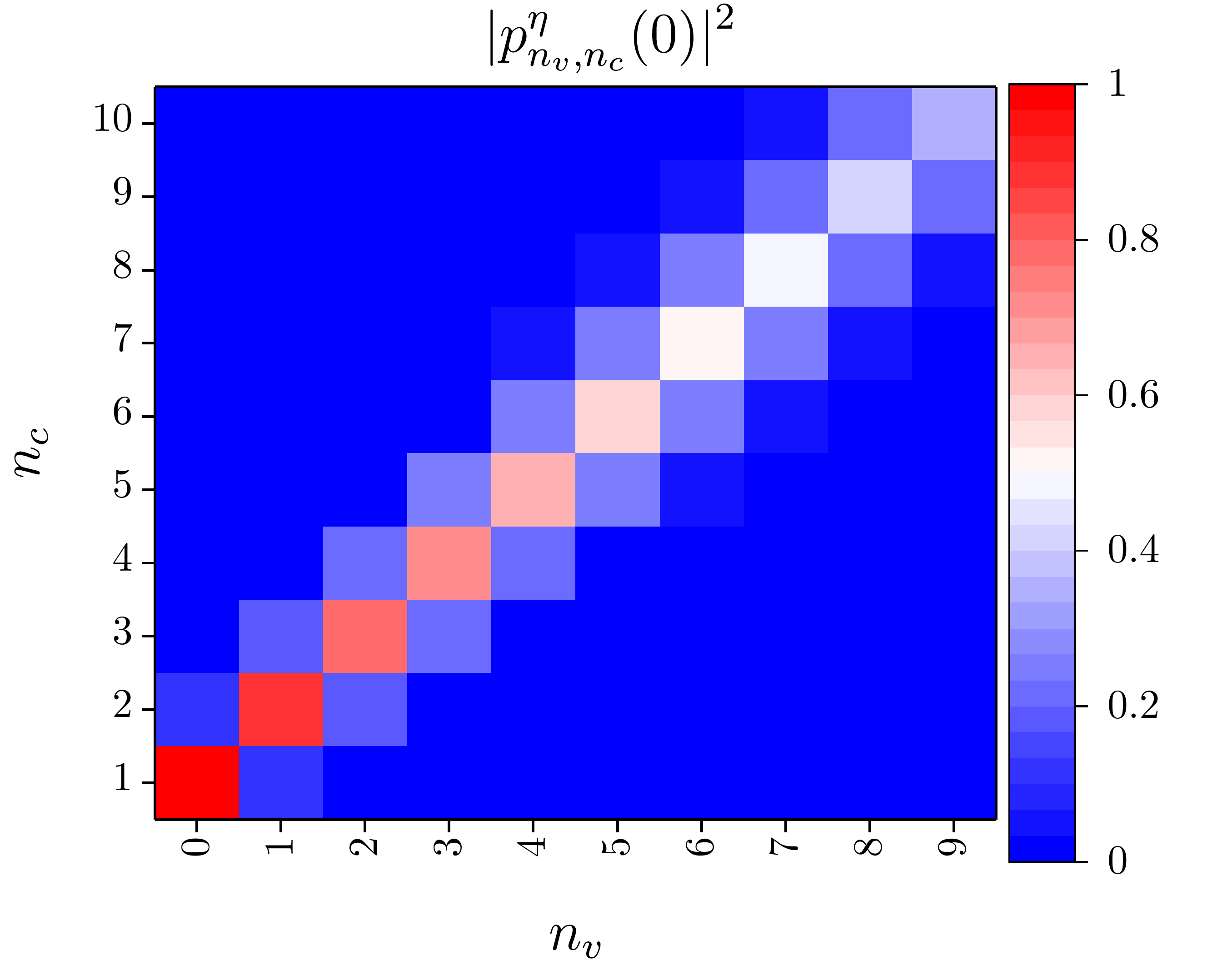}
    \caption{Plot of the squared eigenvector of the $A$ exciton in MoS\textsubscript{2} in an external field of 100 T at $k_y=0$. The elements of the eigenvector have been normalized, such that the largest norm is unity. The plot shows that only a few transitions are significant, and that they are centered around transitions allowed by the optical selection rules.}\label{fig:wavefunc}
\end{figure}

In photoluminescence and spectroscopy experiments, it is typically the exciton transition energy and not the exchange self-energy corrected band gap that is measured. But demonstrating that the exchange self-energy correction is approximately linearly in the magnetic field is important if the diamagnetic shift of the exciton transition energy is used to estimate the exciton size, as was done in Refs.~\onlinecite{stier2016exciton,stier2016probing,stier2018magnetooptics}. Any finite quadratic dependence of the exchange self-energy correction would result in errors in the estimates of the exciton sizes. Although the results presented here do not exclude finite quadratic terms in the exchange self-energy correction, they appear to be small enough that any error in the estimation of the exciton size should be negligible.

\begin{table*}[t]
    \centering
    \begin{tabular}{|c|c|c|c|c|c|c|c|c|c|c|c|c|}
            \toprule
            \multicolumn{2}{|c|}{} & \multicolumn{6}{c|}{Transition energies} & \multicolumn{4}{c|}{Exciton energies}\\ \cline{3-12}
            \multicolumn{2}{|c|}{} & \multicolumn{2}{c|}{EOM} & \multicolumn{2}{c|}{Experimental, $B=0$ T} & \multicolumn{2}{c|}{Experimental, $B \approx 65$ T} & \multicolumn{2}{c|}{EOM} & \multicolumn{2}{c|}{Wannier} \\ \colrule
            TMD & $\kappa$ & $A$ & $B$ & $A$ & $B$ & $A$ & $B$ & $A$ & $B$ & $A$ & $B$ \\ \colrule
            
            MoS\textsubscript{2} & 1.00 & 1.918 & 2.076 &  & & & & -0.620 & -0.632 & -0.617 & -0.632 \\
            {} & 1.55 & 1.907 & 2.066 & 1.895\cite{stier2016exciton}, 1.948\cite{mitioglu2016magnetoexcitons} & 2.042\cite{stier2016exciton}, 2.092\cite{mitioglu2016magnetoexcitons} & 1.896\cite{stier2016exciton}, 1.948\cite{mitioglu2016magnetoexcitons} & 2.044\cite{stier2016exciton}, 2.094\cite{mitioglu2016magnetoexcitons} & -0.491 & -0.504 & -0.489 & -0.503 \\
            
            MoSe\textsubscript{2} & 1.00 & 1.516 & 1.735 & & & & & -0.526 & -0.542 & -0.513 & -0.533 \\
            {} & 1.55 & 1.512 & 1.730 & 1.660\cite{macneill2015breaking}  &  &  &  & -0.419 & -0.434 & -0.409 & -0.428 \\
            
            WS\textsubscript{2} & 1.00 & 2.042 & 2.467 &  & & & & -0.559 & -0.584 & -0.520 & -0.555 \\
            {} & 1.55 & 2.030 & 2.453 & 2.039\cite{stier2016magnetoreflection}, 2.045\cite{stier2016exciton} & 2.442\cite{stier2016magnetoreflection}, 2.453\cite{stier2016exciton} & 2.040\cite{stier2016magnetoreflection}, 2.046\cite{stier2016exciton} & 2.442\cite{stier2016magnetoreflection}, 2.454\cite{stier2016exciton} & -0.426 & -0.450 & -0.392 & -0.424\\

            WSe\textsubscript{2} & 1.00 & 1.761 & 2.216 & &  & &  & -0.511 & -0.535 & -0.468 & -0.505 \\
             & 1.55 & 1.755 & 2.209 & 1.744\cite{aivazian2015magnetic} &  &  &  & -0.393 & -0.417 & -0.357 & -0.391 \\
             & 3.30 & 1.721 & 2.173 & 1.732\cite{stier2016probing} &  & 1.733\cite{stier2016probing} &  & -0.229 & -0.247 & -0.197 & -0.224 \\
             & 4.50 & 1.700 & 2.152 & 1.723\cite{stier2018magnetooptics} &  & 1.724\cite{stier2018magnetooptics} &  & -0.177 & -0.192 & -0.144 & -0.168 \\
            \botrule
     \end{tabular}\caption{Theoretical and experimental transition and exciton energies for $A$ and $B$ excitons in TMDs with different dielectric environments. All theoretical energies are computed at 100 T. Experimental exciton transition energies are indicated by superscripts.}\label{tab:results}
\end{table*}

Turning our attention to the exciton states, we note that it is difficult to separate the bright and dark exciton states calculated in the EOM approach, since Eq.~\eqref{eq:BSE} mixes dark and bright transitions. This difficulty might be resolved by writing the magnetic vector potential in the symmetric gauge in $\hat{H}_B$ and repeating the derivations in Sec.~\ref{sec:electron-electron}, but this study is left for future work. At the present time, we will instead focus on the ground state excitons. We follow convention and denote the spin up and down ground state excitons at the $K$ valley as $A$ and $B$, respectively. Similarly, we have $A'$ and $B'$ ground state excitons in the $K'$ valley. In the absence of valley Zeeman splitting, the $A$ and $A'$ excitons are energetically degenerate and the same holds for the $B$ and $B'$ excitons. Consequently, in the following, only the $A$ and $B$ excitons are considered. In Fig.~\ref{fig:wavefunc}, the squared eigenvector of the $A$ exciton in MoS\textsubscript{2} is plotted for $k_y = 0$. The plot shows that the significant transitions between LLs are where $n_v$ couples to $n_c = n_v + 1$, which coincides exactly the bright transitions according to Sec.~\ref{sec:dipole-elements}. We also find that the same holds for the $B$ exciton. Consequently, the exciton ground states must be bright.

When solving Eq.~\eqref{eq:BSE}, discretizing the integral over $q_y$ using a Gauss-Hermite quadrature with $N_k = 300$ nodes has been found to result in good convergence. If we then include the first $15$ valence and conduction LLs in the summation in Eq.~\eqref{eq:BSE}, the resulting matrix has size $67500 \times 67500$ and is at the limit of what we can handle numerically. But for these values the exciton transition energy has not yet converged, as illustrated for the $A$ exciton in MoS\textsubscript{2} by the black line in Fig.~\ref{fig:convergence}. Alternatively, we can utilize that only a few transitions are significant in the exciton ground state, as was demonstrated in Fig.~\ref{fig:wavefunc}. In fact, calculating the norm of the eigenvector where only transitions of the type $n_v$ to $n_c \in [n_v-1, n_v+3]$, have been included, we find that the squared overlap is only 2\% less than unity. Including only these significant transitions allows us to include more valence LLs and, as illustrated by the red line in Fig.~\ref{fig:convergence}, obtain a better convergence. The cost is a small error on the order of a few meV. The numerical difficulties associated with including a high number of LLs in the excitonic calculations result in a restriction on the magnetic field strength used hence, as the magnetic field strength decreases, more LLs need to be included in the calculations to secure sufficiently converged results. Eventually, the current computational restrictions limit us to magnetic fields above 100 T.

Turning to the exciton transition energies, we begin by expressing the transition energies $E_{\tau}$ in terms of the different magnetic field-dependent terms. At low magnetic fields, we can write\cite{stier2016exciton,zipfel2018spatial}
\begin{equation}
E_{\tau} = E_0 + \mu_g B + \tau \mu_Z B + \sigma_{dia} B^2,
\end{equation}
with $\tau$ the valley index, $E_0$ the zero-field exciton transition energy, $\mu_gB$  the field dependent change in band gap, $\tau \mu_Z B$ the valley Zeeman shift, and finally $\sigma_{dia} B^2$ the diamagnetic shift. Since the valley Zeeman shift is not included in our single-particle Hamiltonian, the transition energies found by solving Eq.~\eqref{eq:BSE} are of the form $E = E_0 + \mu_g B + \sigma_{dia} B^2$. To allow for comparisons between the theoretical and the experimentally measured exciton transitions energies, we average the experimentally measured exciton transition energies from the $K$ and $K'$ valleys to remove the valley Zeeman splitting, i.e. use $E = (E_{+1}+E_{-1})/2$.

The exciton transition energies of the $A$ and $B$ excitons are presented in Table~\ref{tab:results}. In columns three and four, we show the theoretical transition energies, which were calculated by solving Eq.~\eqref{eq:BSE}. Columns five and six, contain the experimental exciton transition energies when there is no external magnetic field. In columns seven and eight, we show the experimental exciton transition energies at approximately $65$ T. Comparing the zero-field transition energies with the experimental transition energies in columns seven and eight, we see that the exciton transition energies exhibit a minimal dependence on the magnetic field. In fact, experiments predict that the quadratic diamagnetic shift is on the order of only a few meV\cite{stier2016exciton,stier2018magnetooptics} for a magnetic field of 100 T. Consequently, we can compare the calculated transition energies to the measured transition energies in a system with no magnetic field. Table~\ref{tab:results} shows that the transition energies of MoS\textsubscript{2}, WS\textsubscript{2}, and WSe\textsubscript{2} are very well captured by our model, with differences on the order of 10 meV. The calculated results for MoSe\textsubscript{2} differ more from the experimental results, with the calculated transition energy being approximately 150 meV below the experimental transition energy. This discrepancy indicates a problem with the material parameters used and not the method, as the results agree well for the three other types of materials.

In the final four columns of Table~\ref{tab:results}, the exciton energies calculated using the EOM approach and the Wannier model are presented. For the EOM method, the exciton energies are found from $E_{exc} = E-\tilde{E}_g$, where $E$ is the exciton transition energy found by solving Eq.~\eqref{eq:BSE} and $\tilde{E}_g$ is the exchange self-energy corrected band gap. Comparing the results, we see that all the exciton energies calculated using the EOM approach are below the Wannier results. That is to be expected since the EOM approach relies on less strict approximations. The differences between the calculated energies are quite small and vary from a few meV to 50 meV. Thus, if errors in this range are acceptable, the Wannier model provides a useful model for excitons in monolayer TMDs.

\begin{figure}[t]
    \centering
    \includegraphics[width=0.48\textwidth]{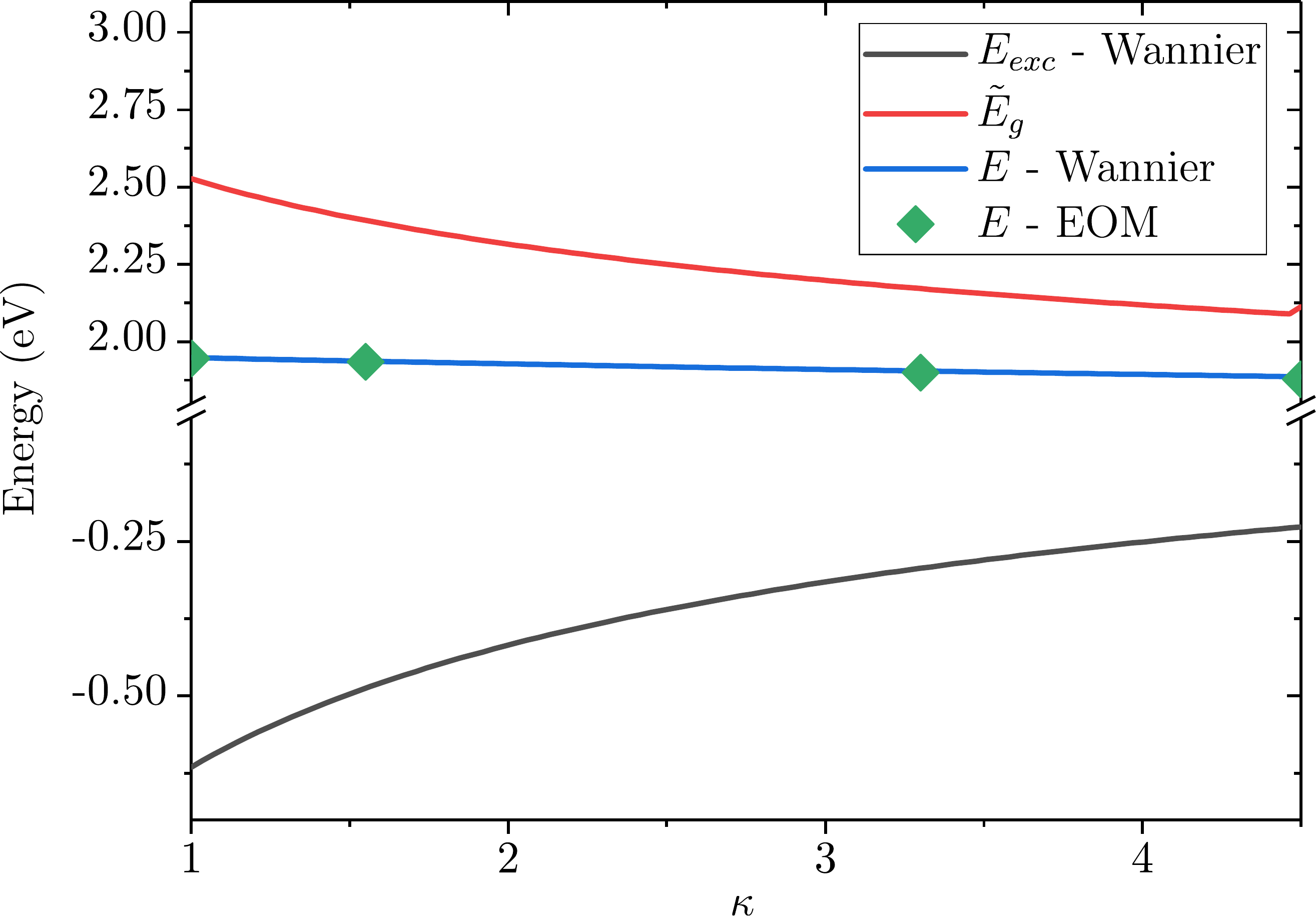}
    \caption{Plot of the corrected band gap (red line), the exciton transition energy (blue line and green diamonds) and the exciton energy (black line) as a function of the relative dielectric constant of the surrounding medium for MoS\textsubscript{2}, with $B=100$ T and $\tau s = +1$. The exciton transition energy calculated from the Wannier results (blue line) is the sum of exciton energy (black line) and the corrected band gap (red line), i.e. $E = \tilde{E}_g+E_{exc}$.}\label{fig:renorm}
\end{figure}

Finally, we also consider the effect of changing the dielectric environment of the TMDs, i.e. varying the screening parameter $\kappa$ in the potentials in Eqs.~\eqref{eq:keldysh} and \eqref{eq:keldysh_real}. The effect is illustrated in Fig.~\ref{fig:renorm} for MoS\textsubscript{2} in a magnetic field of 100 T. The figure shows that the exchange self-energy corrected band gap decreases while the exciton energy increases as a function of $\kappa$. These two counteracting effects result in exciton transition energies, which only exhibit minimal dependence on the dielectric environment, as illustrated by the blue line and green squares in Fig.~\ref{fig:renorm}. This effect has previously been demonstrated in TMDs with no external magnetic field\cite{lin2014dielectric}, but Fig.~\ref{fig:renorm} illustrates that it still holds for systems in the presence of a perpendicular magnetic field. This phenomenon further underlines the importance of including the exchange self-energy corrections in a self-contained model. We find that similar results hold for the other TMDs.

Comparing the EOM method and the Wannier model, we see that both have advantages and disadvantages. The EOM method provides a self-contained framework, including the unique LL structure and a higher accuracy of the exciton energies. The disadvantage is that the numerical computations are demanding and, as a consequence, small magnetic fields cannot be considered. For the Wannier method, the numerical calculations are relatively simple and arbitrary magnetic field strengths can be considered. The disadvantages are that for some systems the accuracy is lower than the EOM method and that only the excitonic properties are described. The Wannier model provides no information about the unique LL structure or the field-dependent change of the band gap. Consequently, the choice between the EOM method and the Wannier method depends on the application, and which aspects are deemed important.

\section{Summary}
\label{sec:summary}
In summary, starting from a Dirac-type Hamiltonian describing the band structure of monolayer TMDs around the $K$ and $K'$ points, we have introduced an external magnetic field and then included electron-electron interactions to account for the exchange self-energy corrections and excitons. In this setup, we used the EOM approach to find the ground state exciton transition energies. Our results were compared to the popular Wannier model for excitons and recent experimental results. 

When comparing with the Wannier model, we found that the ground state exciton energies match quite well. Consequently, the EOM method validates the Wannier model in this case. The exciton energies only exhibit a small dependence on the magnetic field (up to a few meV for realistic field strengths), but the optical properties are expected to change significantly. Thus, we will focus on the optical properties of magnetoexcitons in future projects. We also expect to see more pronounced differences between the optical response calculated using the EOM approach and the Wannier model. 

Comparing the calculated transition energies with the experimental values, we also found a very good agreement. This shows that the exchange self-energy correction is central if accurate theoretical calculations of the exciton transition energies are needed. Finally, we considered the effect of the dielectric environment on the exciton transition energy. We found that increasing the dielectric constant of the environment causes a decrease in the corrected band gap and an increase in the exciton energy. These two counteracting effects cause a minimal dependence of the exciton transition energies on the dielectric environment. This holds for both the EOM method results and transition energies calculated from the Wannier model results.

\section*{Acknowledgments}
J.H. and T.G.P. gratefully acknowledges financial support by the QUSCOPE Center, sponsored by the Villum Foundation. Additionally, T.G.P. is supported by the Center for Nanostructured Graphene (CNG), which is sponsored by the Danish National Research Foundation, Project No. DNRF103. G. C. acknowledges financial support from FCT for the P2020-PTDC/FIS-NAN/4662/2014 project. N.M.R.P. acknowledges support from the European Commission through the project “Graphene- Driven Revolutions in ICT and Beyond" (Ref. No. 785219), and the Portuguese Foundation for Science and Technology (FCT) in the framework of the Strategic Financing UID/FIS/04650/2013. Additionally, N.M.R.P. acknowledges  COMPETE2020, PORTUGAL2020, FEDER and the Portuguese Foundation for Science and Technology (FCT) through project PTDC/FIS-NAN/3668/201

\appendix
\begin{widetext}
\section{Commutator relations and the equation of motion}\label{app:commutators}
In this section, we present the commutator relations between $\hat{H} = \hat{H}_B +\hat{H}_I +\hat{H}_{ee}$ and the density matrix, as well as the relevant equation of motion. First, we calculate the commutator relations using the following relation
\begin{equation}
    \left[\hat{\rho}^\eta_{\alpha_1,\alpha_2},\hat{\rho}^{\eta'}_{\alpha_3,\alpha_4}\right] = \hat{\rho}_{\alpha_1,\alpha_4}^\eta\delta_{\alpha_2,\alpha_3}\delta_{\eta,\eta'} - \hat{\rho}_{\alpha_3,\alpha_2}^\eta\delta_{\alpha_1,\alpha_4}\delta_{\eta,\eta'}.
\end{equation}
Applying this relation to the first two terms of the commutator $[\hat{H}, \rho^\eta_{\alpha,\alpha'}]$, we find
\begin{align}
    \left[\hat{H}_B,\hat{\rho}^{\eta}_{\alpha,\alpha'}\right] &= \sum_{\alpha'',\eta'}E^{\eta'}_{\alpha''}\left[\hat{\rho}^{\eta'}_{\alpha'',\alpha''},\hat{\rho}^{\eta}_{\alpha,\alpha'}\right] \\
    & = (E^\eta_\alpha - E^\eta_{\alpha'}) \hat{\rho}^\eta_{\alpha,\alpha'}\label{eq:commutator_HB},
\end{align}
and
\begin{align}
    \left[\hat{H}_I,\hat{\rho}^{\eta}_{\alpha,\alpha'}\right] &=  -\bm{\mathcal{E}}(t)\cdot\sum_{\alpha_1,\alpha_2,\eta'}  \mathbf{d}_{\eta'}^{\alpha_1 \to \alpha_2}\left[\hat{\rho}^{\eta'}_{\alpha_1,\alpha_2},\hat{\rho}^{\eta}_{\alpha,\alpha'}\right] \\
   &=  -\bm{\mathcal{E}}(t)\cdot\sum_{\alpha''} \left(  \mathbf{d}_\eta^{\alpha'' \to \alpha}\hat{\rho}_{\alpha'',\alpha'}^\eta - \mathbf{d}_\eta^{\alpha' \to \alpha''}\hat{\rho}_{\alpha,\alpha''}^\eta\right).\label{eq:commutator_HI}
\end{align}
In the commutator relation between the electron-electron interaction Hamiltonian and the density matrix, the following commutator relation is useful
\begin{align}
\left[\hat{c}^\dagger_{\alpha_1,\tau,s'}\hat{c}^\dagger_{\alpha_2,\tau,s''}\hat{c}_{\alpha_3,\tau,s''}\hat{c}_{\alpha_4,\tau,s'}, \hat{c}^\dagger_{\alpha, \tau', s}\hat{c}_{\alpha', \tau', s}\right] = \delta_{\tau,\tau'} & \left(\hat{c}^\dagger_{\alpha_1,\tau,s}\hat{c}^\dagger_{\alpha_2,\tau,s''}\hat{c}_{\alpha_3,\tau,s''}\hat{c}_{\alpha',\tau,s}\delta_{\alpha,\alpha_4}\delta_{s,s'}\right. \nonumber \\
& \quad +\hat{c}^\dagger_{\alpha_1,\tau,s'}\hat{c}^\dagger_{\alpha_2,\tau,s}\hat{c}_{\alpha',\tau,s}\hat{c}_{\alpha_4,\tau,s'}\delta_{\alpha,\alpha_3}\delta_{s,s''} \nonumber \\
& \quad -\hat{c}^\dagger_{\alpha,\tau,s}\hat{c}^\dagger_{\alpha_2,\tau,s''}\hat{c}_{\alpha_3,\tau,s''}\hat{c}_{\alpha_4,\tau,s}\delta_{\alpha',\alpha_1}\delta_{s,s'} \nonumber \\
& \left. \quad -\hat{c}^\dagger_{\alpha_1,\tau,s'}\hat{c}^\dagger_{\alpha,\tau,s}\hat{c}_{\alpha_3,\tau,s}\hat{c}_{\alpha_4,\tau,s'}\delta_{\alpha',\alpha_2}\delta_{s,s''}\right).\label{eq:commutator}
\end{align}
Applying Eq.~\eqref{eq:commutator} to the $[\hat{H}_{ee}, \hat{\rho}^\eta_{\alpha,\alpha'}]$ commutator, we find
\begin{equation}\label{eq:ee-commutator}
\left[H_{ee}, \hat{\rho}_{\alpha,\alpha'}^{\eta}\right] = \sum_{\substack{s',\alpha_1 \\ \alpha_2, \alpha_3}}\left\{U^{\tau, s, s'}_{\alpha_1 \alpha, \alpha_2 \alpha_3}\hat{c}^\dagger_{\alpha_1,\tau,s}\hat{c}^\dagger_{\alpha_2,\tau,s'}\hat{c}_{\alpha_3,\tau,s'}\hat{c}_{\alpha',\tau,s}- U^{\tau, s, s'}_{\alpha' \alpha_1, \alpha_2 \alpha_3}\hat{c}^\dagger_{\alpha,\tau,s}\hat{c}^\dagger_{\alpha_2,\tau,s'}\hat{c}_{\alpha_3,\tau,s'}\hat{c}_{\alpha_1,\tau,s}\right\},
\end{equation}
where we also used the relation
\begin{equation}
U^{\tau, s, s'}_{\alpha_1 \alpha_4, \alpha_2 \alpha_3} = U^{\tau, s', s}_{\alpha_2 \alpha_3, \alpha_1 \alpha_4}.
\end{equation}
Collecting the terms in Eqs.~\eqref{eq:commutator_HB}, \eqref{eq:commutator_HI} and \eqref{eq:ee-commutator}, we can now write Heisenberg's equation of motion for the full Hamiltonian including electron-electron interactions. To write Eq.~\eqref{eq:eos-exp}, we compute the expectation value of the commutator relations keeping terms, which are of first order in the electric field. While the expectation values of Eqs.~\eqref{eq:commutator_HB} and \eqref{eq:commutator_HI} are found by straightforward calculation, we apply the random phase approximation (RPA)\cite{ehrenreich1959self} to find
\begin{align}
    \left\langle\left[H_{ee}, \hat{\rho}_{\alpha,\alpha'}^{\tau,s}\right]\right\rangle = \sum_{\substack{s',\alpha_1 \\ \alpha_2, \alpha_3}}&\left\{U^{\tau, s, s'}_{\alpha_1 \alpha, \alpha_2 \alpha_3}\left(p_{\alpha_2,\alpha_3}^{\tau, s'}p_{\alpha_1,\alpha'}^{\tau, s} - \delta_{s,s'}p_{\alpha_1,\alpha_3}^{\tau, s}p_{\alpha_2,\alpha'}^{\tau, s}\right)\right. \nonumber \\ 
    & \quad - \left. U^{\tau, s, s'}_{\alpha' \alpha_1, \alpha_2 \alpha_3}\left(p_{\alpha_2,\alpha_3}^{\tau, s'}p_{\alpha,\alpha_1}^{\tau, s}-\delta_{s,s'}p_{\alpha_2,\alpha_1}^{\tau, s}p_{\alpha,\alpha_3}^{\tau, s}\right)\right\},\label{eq:rpa-commutator}
\end{align}
where $p_{\alpha,\alpha'}^{\tau, s} = \langle\hat{\rho}_{\alpha,\alpha'}^{\tau, s}\rangle$. Terms allowing mixing of spins correspond to the Hartree terms in Hartree-Fock theory. They are canceled by the interaction with the positive background\cite{mahan2013many} and, as a result, the expectation value has the following form
\begin{equation}
    \left\langle\left[H_{ee}, \hat{\rho}_{\alpha,\alpha'}^{\tau,s}\right]\right\rangle =\sum_{\alpha_1,\alpha_3} p_{\alpha_1, \alpha_3}\sum_{\substack{\alpha_2}} \left( U^{\tau, s, s}_{\alpha' \alpha_3, \alpha_1 \alpha_2}p_{\alpha,\alpha_2}^{\tau, s} - U^{\tau, s, s}_{\alpha_1 \alpha, \alpha_2 \alpha_3}p_{\alpha_2,\alpha'}^{\tau, s}\right).\label{eq:rpa-commutator}
\end{equation}
This gives the following EOM for the expectation value
\begin{equation}\label{eq:complete-eom-app}
    \left(E_{\alpha'}^\eta-E_{\alpha}^\eta-i\hbar\frac{\partial }{\partial t}\right) p^\eta_{\alpha, \alpha'} =   \sum_{\substack{\alpha_1,\alpha_2 \\ \alpha_3}} p_{\alpha_1, \alpha_3}^\eta \left( U^{\tau, s, s}_{\alpha' \alpha_3, \alpha_1 \alpha_2}p_{\alpha,\alpha_2}^\eta - U^{\tau, s, s}_{\alpha_1 \alpha, \alpha_2 \alpha_3}p_{\alpha_2,\alpha'}^\eta\right)-\bm{\mathcal{E}}(t)\cdot\sum_{\alpha''} \left(  \mathbf{d}_\eta^{\alpha'' \to \alpha}p_{\alpha'',\alpha'}^\eta - \mathbf{d}_\eta^{\alpha' \to \alpha''}p_{\alpha,\alpha''}^\eta\right).
\end{equation}
The final step is to expand the expectation values in orders of the electric field and collect first-order terms in Eq.~\eqref{eq:complete-eom-app}. The zero'th order of the expectation value can be expressed using the Fermi-Dirac distribution
\begin{equation}
    p_{\alpha,\alpha'}^{\eta,0} = f(E^\eta_\alpha)\delta_{\alpha,\alpha'},
\end{equation}
where $f(E)$ is the Fermi-Dirac distribution. Consequently, the first order equation is 
\begin{align}
    \left(E^\eta_{\alpha'}-E^\eta_\alpha -i\hbar\frac{\partial }{\partial t}\right) p^{\eta,1}_{\alpha, \alpha'} &= \left(\sum_{\alpha_1,\alpha_2}U_{\alpha' \alpha_2, \alpha_1 \alpha}^{\tau, s,s}p_{\alpha_1,\alpha_2}^{\eta, 1}-\bm{\mathcal{E}}(t)\cdot \mathbf{d}_{\eta}^{\alpha' \to \alpha}\right)\Delta f_{\alpha',\alpha}^{\eta}\nonumber \\
     & \quad +\sum_{\alpha_1,\alpha_2} f(E_{\alpha_1}^\eta) \left( U^{\tau, s, s}_{\alpha' \alpha_1, \alpha_1, \alpha_2}p_{\alpha,\alpha_2}^{\eta, 1} - U^{\tau, s, s}_{\alpha_1 \alpha, \alpha_2 \alpha_1}p_{\alpha_2,\alpha'}^{\eta, 1}\right),
\end{align}
where $\Delta f_{\alpha',\alpha}^{\eta} =  f(E_{\alpha'}^\eta)-f(E_\alpha^\eta)$ and $p^{\eta,1}_{\alpha, \alpha'}$ is the first-order term of the expectation value. We rewrite the last term on the right hand side to isolate the exchange self-energy correction
\begin{align}\label{eq:density}
    \sum_{\alpha_1,\alpha_2} f(E_{\alpha_1}^\eta) \left( U^{\tau, s, s}_{\alpha' \alpha_1, \alpha_1, \alpha_2}p_{\alpha,\alpha_2}^{\eta, 1} - U^{\tau, s, s}_{\alpha_1 \alpha, \alpha_2 \alpha_1}p_{\alpha_2,\alpha'}^{\eta, 1}\right) &= \Sigma_{\alpha'}^\eta - \Sigma_{\alpha}^\eta + \sum_{\alpha_1}f(E_{\alpha_1}^\eta) \times \nonumber \\
    &\left(\sum_{\alpha_2 \neq \alpha'}U^{\tau, s, s}_{\alpha' \alpha_1, \alpha_1 \alpha_2}p_{\alpha,\alpha_2}^{\eta, 1}-\sum_{\alpha_2 \neq \alpha}U^{\tau, s, s}_{\alpha_1 \alpha, \alpha_2 \alpha_1}p_{\alpha_2,\alpha'}^{\eta, 1}\right),
\end{align}
where $\Sigma_{\alpha}^\eta$ is the exchange self-energy correction given by
\begin{equation}\label{eq:self-energy}
    \Sigma_\alpha^\eta = \sum_{\alpha_1}f(E_{\alpha_1}^\eta) U_{\alpha_1 \alpha, \alpha \alpha_1}^{\tau, s, s}.
\end{equation}
The remaining terms in Eq.~\eqref{eq:density} correspond to density terms and will be disregarded in this work. Thus, the first order EOM for the expectation value of the density matrix reads
\begin{equation}\label{eq:final-eom}
    \left(\tilde{E}^\eta_{\alpha'}-\tilde{E}^\eta_\alpha -i\hbar\frac{\partial }{\partial t}\right) p^{\eta,1}_{\alpha, \alpha'} = \left(\sum_{\alpha_1,\alpha_2}U_{\alpha' \alpha_2, \alpha_1 \alpha}^{\tau, s,s}p_{\alpha_1,\alpha_2}^{\eta, 1}-\bm{\mathcal{E}}(t)\cdot \mathbf{d}_{\eta}^{\alpha' \to \alpha}\right)\Delta f_{\alpha',\alpha}^{\eta},
\end{equation}
with $\tilde{E}_\alpha^\eta = E_\alpha^\eta - \Sigma_\alpha^\eta$. The interband solutions to the system of first order differential equations in Eq.~\eqref{eq:final-eom} give the excitonic states. 

\section{Structure factors}\label{app:formfactors}
In this section, we find an explicit expression for the structure factors $F^{\tau, s}_{\alpha,\alpha'}$ defined in Eq.~\eqref{eq:structure-factors}. The explicit expression allows for a numerical evaluation of the Coulomb integrals in Eq.~\eqref{eq:coulomb-integrals}. Inserting the expression for the single-particle wavefunction, Eq.~\eqref{eq:ss-wavefunction}, in the structure factors, we find
\begin{equation}\label{eq:structure-factor2}
    F_{\alpha,\alpha'}^{\tau, s} = \int \mathrm{d}^2\mathbf{r} ~\frac{e^{i(q_y-k_y+k_y')y}}{L_y} e^{iq_x x}\left(B^{n,\lambda}_{\tau, s}B^{n',\lambda'}_{\tau, s}\phi_{n_{\tau,-}}(\tilde{x})\phi_{n_{\tau,-}'}(\tilde{x}') + C^{n,\lambda}_{\tau, s}C^{n',\lambda'}_{\tau, s}\phi_{n_{\tau,+}}(\tilde{x})\phi_{n_{\tau,+}'}(\tilde{x}')\right),
\end{equation}
where the notation is $\tilde{x} = x + l_B^2 k_y$, $\tilde{x}' = x + l_B^2k_y'$, $n_{\tau, -} = n-(\tau+1)/2 $ and $n_{\tau, +} = n+(\tau-1)/2$. For each term of Eq.~\eqref{eq:structure-factor2}, we calculate an integral of the type
\begin{align}\label{eq:details}
    \int \mathrm{d}x ~ e^{iq_x x}\phi_{n}(\tilde{x})\phi_{n'}(\tilde{x}') &= \exp\left(-\frac{l_B^2(k_y-k_y')^2+l_B^2q_x^2}{4}+i q_x\frac{l_B^2}{2}(k_y+k_y')\right) \nonumber \\ 
    &\quad \times \sqrt{\frac{n_{<}!}{n_{>}!}}\left(\frac{il_B q_x+l_B\sgn(n-n')(k_y-k_y')}{\sqrt{2}}\right)^{n_{>}-n_{<}}L_{n_{<}}^{n_{>}-n_{<}}\left(\frac{l_B^2q^2_x+l_B^2(k_y-k_y')^2}{2}\right),
\end{align}
where $n_{>} = \max\{n,n'\}$, $n_{<} = \min\{n, n'\}$ and $L_n^m$ are associated Laguerre polynomials. The detailed calculation of the integral in Eq.~\eqref{eq:details} was provided in Ref.~\onlinecite{bychkov1983two}. The previous expression allows us to write the structure factors as
\begin{align}\label{eq:final-structure}
    F^{\tau, s}_{\alpha,\alpha'}(\mathbf{q}) = \frac{\pi \delta(q_y-k_y+k_y')}{L_y}\exp\left(-\frac{l_B^2 |\mathbf{q}|^2}{4}+i q_x\frac{l_B^2}{2}(k_y+k_y')\right)J_{\lambda n,\lambda' n'}^{\tau, s}(\mathbf{q}),
\end{align}
where the function $J_{\lambda n, \lambda' n'}^\eta$ is defined as
\begin{align}
    J_{\lambda n,\lambda' n'}^{\tau, s}(\mathbf{q})=&\left(\frac{il_Bq_x+l_B\sgn(n-n')q_y}{\sqrt{2}}\right)^{n_{>}-n_{<}}\left(\sqrt{\frac{(n_{<}-(\tau+1)/2)!}{(n_{>}-(\tau+1)/2)!}}B^{n,\lambda}_{\tau, s}B^{n',\lambda'}_{\tau, s}L_{n_{<}-(\tau+1)/2}^{n_{>}-n_{<}}\left(\frac{l_B^2|\mathbf{q}|^2}{2}\right)\right. \nonumber\\ &\quad \left. + \sqrt{\frac{(n_{<}+(\tau-1)/2)!}{(n_{>}+(\tau-1)/2)!}}C^{n,\lambda}_{\tau, s}C^{n',\lambda'}_{\tau, s}L_{n_{<}+(\tau-1)/2}^{n_{>}-n_{<}}\left(\frac{l_B^2 |\mathbf{q}|^2}{2}\right)\right).\label{eq:J_function}
\end{align}
The expression for the structure factors in Eq.~\eqref{eq:final-structure} is used to calculate both the excitonic properties and the exchange self-energy corrections.
\end{widetext}
\bibliographystyle{aipnum4-1}
\bibliography{references}

\begin{thebibliography}{51}%
\makeatletter
\providecommand \@ifxundefined [1]{%
 \@ifx{#1\undefined}
}%
\providecommand \@ifnum [1]{%
 \ifnum #1\expandafter \@firstoftwo
 \else \expandafter \@secondoftwo
 \fi
}%
\providecommand \@ifx [1]{%
 \ifx #1\expandafter \@firstoftwo
 \else \expandafter \@secondoftwo
 \fi
}%
\providecommand \natexlab [1]{#1}%
\providecommand \enquote  [1]{``#1''}%
\providecommand \bibnamefont  [1]{#1}%
\providecommand \bibfnamefont [1]{#1}%
\providecommand \citenamefont [1]{#1}%
\providecommand \href@noop [0]{\@secondoftwo}%
\providecommand \href [0]{\begingroup \@sanitize@url \@href}%
\providecommand \@href[1]{\@@startlink{#1}\@@href}%
\providecommand \@@href[1]{\endgroup#1\@@endlink}%
\providecommand \@sanitize@url [0]{\catcode `\\12\catcode `\$12\catcode
  `\&12\catcode `\#12\catcode `\^12\catcode `\_12\catcode `\%12\relax}%
\providecommand \@@startlink[1]{}%
\providecommand \@@endlink[0]{}%
\providecommand \url  [0]{\begingroup\@sanitize@url \@url }%
\providecommand \@url [1]{\endgroup\@href {#1}{\urlprefix }}%
\providecommand \urlprefix  [0]{URL }%
\providecommand \Eprint [0]{\href }%
\providecommand \doibase [0]{http://dx.doi.org/}%
\providecommand \selectlanguage [0]{\@gobble}%
\providecommand \bibinfo  [0]{\@secondoftwo}%
\providecommand \bibfield  [0]{\@secondoftwo}%
\providecommand \translation [1]{[#1]}%
\providecommand \BibitemOpen [0]{}%
\providecommand \bibitemStop [0]{}%
\providecommand \bibitemNoStop [0]{.\EOS\space}%
\providecommand \EOS [0]{\spacefactor3000\relax}%
\providecommand \BibitemShut  [1]{\csname bibitem#1\endcsname}%
\let\auto@bib@innerbib\@empty
\bibitem [{\citenamefont {Tanaka}, \citenamefont {Fukutani},\ and\
  \citenamefont {Kuwabara}(1978)}]{tanaka1978excitons}%
  \BibitemOpen
  \bibfield  {author} {\bibinfo {author} {\bibfnamefont {M.}~\bibnamefont
  {Tanaka}}, \bibinfo {author} {\bibfnamefont {H.}~\bibnamefont {Fukutani}}, \
  and\ \bibinfo {author} {\bibfnamefont {G.}~\bibnamefont {Kuwabara}},\
  }\href@noop {} {\bibfield  {journal} {\bibinfo  {journal} {J. Phys. Soc.
  Jpn.}\ }\textbf {\bibinfo {volume} {45}},\ \bibinfo {pages} {1899} (\bibinfo
  {year} {1978})}\BibitemShut {NoStop}%
\bibitem [{\citenamefont {Xiao}\ \emph {et~al.}(2012)\citenamefont {Xiao},
  \citenamefont {Liu}, \citenamefont {Feng}, \citenamefont {Xu},\ and\
  \citenamefont {Yao}}]{xiao2012coupled}%
  \BibitemOpen
  \bibfield  {author} {\bibinfo {author} {\bibfnamefont {D.}~\bibnamefont
  {Xiao}}, \bibinfo {author} {\bibfnamefont {G.-B.}\ \bibnamefont {Liu}},
  \bibinfo {author} {\bibfnamefont {W.}~\bibnamefont {Feng}}, \bibinfo {author}
  {\bibfnamefont {X.}~\bibnamefont {Xu}}, \ and\ \bibinfo {author}
  {\bibfnamefont {W.}~\bibnamefont {Yao}},\ }\href@noop {} {\bibfield
  {journal} {\bibinfo  {journal} {Phys. Rev. Lett.}\ }\textbf {\bibinfo
  {volume} {108}},\ \bibinfo {pages} {196802} (\bibinfo {year}
  {2012})}\BibitemShut {NoStop}%
\bibitem [{\citenamefont {Korm{\'a}nyos}\ \emph {et~al.}(2013)\citenamefont
  {Korm{\'a}nyos}, \citenamefont {Z{\'o}lyomi}, \citenamefont {Drummond},
  \citenamefont {Rakyta}, \citenamefont {Burkard},\ and\ \citenamefont
  {Fal'ko}}]{kormanyos2013monolayer}%
  \BibitemOpen
  \bibfield  {author} {\bibinfo {author} {\bibfnamefont {A.}~\bibnamefont
  {Korm{\'a}nyos}}, \bibinfo {author} {\bibfnamefont {V.}~\bibnamefont
  {Z{\'o}lyomi}}, \bibinfo {author} {\bibfnamefont {N.~D.}\ \bibnamefont
  {Drummond}}, \bibinfo {author} {\bibfnamefont {P.}~\bibnamefont {Rakyta}},
  \bibinfo {author} {\bibfnamefont {G.}~\bibnamefont {Burkard}}, \ and\
  \bibinfo {author} {\bibfnamefont {V.~I.}\ \bibnamefont {Fal'ko}},\
  }\href@noop {} {\bibfield  {journal} {\bibinfo  {journal} {Phys. Rev. B}\
  }\textbf {\bibinfo {volume} {88}},\ \bibinfo {pages} {045416} (\bibinfo
  {year} {2013})}\BibitemShut {NoStop}%
\bibitem [{\citenamefont {Korm{\'a}nyos}\ \emph {et~al.}(2015)\citenamefont
  {Korm{\'a}nyos}, \citenamefont {Burkard}, \citenamefont {Gmitra},
  \citenamefont {Fabian}, \citenamefont {Z{\'o}lyomi}, \citenamefont
  {Drummond},\ and\ \citenamefont {Fal’ko}}]{kormanyos2015k}%
  \BibitemOpen
  \bibfield  {author} {\bibinfo {author} {\bibfnamefont {A.}~\bibnamefont
  {Korm{\'a}nyos}}, \bibinfo {author} {\bibfnamefont {G.}~\bibnamefont
  {Burkard}}, \bibinfo {author} {\bibfnamefont {M.}~\bibnamefont {Gmitra}},
  \bibinfo {author} {\bibfnamefont {J.}~\bibnamefont {Fabian}}, \bibinfo
  {author} {\bibfnamefont {V.}~\bibnamefont {Z{\'o}lyomi}}, \bibinfo {author}
  {\bibfnamefont {N.~D.}\ \bibnamefont {Drummond}}, \ and\ \bibinfo {author}
  {\bibfnamefont {V.}~\bibnamefont {Fal’ko}},\ }\href@noop {} {\bibfield
  {journal} {\bibinfo  {journal} {2D Materials}\ }\textbf {\bibinfo {volume}
  {2}},\ \bibinfo {pages} {022001} (\bibinfo {year} {2015})}\BibitemShut
  {NoStop}%
\bibitem [{\citenamefont {Ramasubramaniam}(2012)}]{ramasubramaniam2012large}%
  \BibitemOpen
  \bibfield  {author} {\bibinfo {author} {\bibfnamefont {A.}~\bibnamefont
  {Ramasubramaniam}},\ }\href@noop {} {\bibfield  {journal} {\bibinfo
  {journal} {Phys. Rev. B}\ }\textbf {\bibinfo {volume} {86}},\ \bibinfo
  {pages} {115409} (\bibinfo {year} {2012})}\BibitemShut {NoStop}%
\bibitem [{\citenamefont {Berkelbach}, \citenamefont {Hybertsen},\ and\
  \citenamefont {Reichman}(2013)}]{berkelbach2013theory}%
  \BibitemOpen
  \bibfield  {author} {\bibinfo {author} {\bibfnamefont {T.~C.}\ \bibnamefont
  {Berkelbach}}, \bibinfo {author} {\bibfnamefont {M.~S.}\ \bibnamefont
  {Hybertsen}}, \ and\ \bibinfo {author} {\bibfnamefont {D.~R.}\ \bibnamefont
  {Reichman}},\ }\href@noop {} {\bibfield  {journal} {\bibinfo  {journal}
  {Phys. Rev. B}\ }\textbf {\bibinfo {volume} {88}},\ \bibinfo {pages} {045318}
  (\bibinfo {year} {2013})}\BibitemShut {NoStop}%
\bibitem [{\citenamefont {Chaves}\ \emph {et~al.}(2017)\citenamefont {Chaves},
  \citenamefont {Ribeiro}, \citenamefont {Frederico},\ and\ \citenamefont
  {Peres}}]{chaves2017excitonic}%
  \BibitemOpen
  \bibfield  {author} {\bibinfo {author} {\bibfnamefont {A.}~\bibnamefont
  {Chaves}}, \bibinfo {author} {\bibfnamefont {R.}~\bibnamefont {Ribeiro}},
  \bibinfo {author} {\bibfnamefont {T.}~\bibnamefont {Frederico}}, \ and\
  \bibinfo {author} {\bibfnamefont {N.}~\bibnamefont {Peres}},\ }\href@noop {}
  {\bibfield  {journal} {\bibinfo  {journal} {2D Materials}\ }\textbf {\bibinfo
  {volume} {4}},\ \bibinfo {pages} {025086} (\bibinfo {year}
  {2017})}\BibitemShut {NoStop}%
\bibitem [{\citenamefont {Rose}, \citenamefont {Goerbig},\ and\ \citenamefont
  {Pi{\'e}chon}(2013)}]{rose2013spin}%
  \BibitemOpen
  \bibfield  {author} {\bibinfo {author} {\bibfnamefont {F.}~\bibnamefont
  {Rose}}, \bibinfo {author} {\bibfnamefont {M.}~\bibnamefont {Goerbig}}, \
  and\ \bibinfo {author} {\bibfnamefont {F.}~\bibnamefont {Pi{\'e}chon}},\
  }\href@noop {} {\bibfield  {journal} {\bibinfo  {journal} {Phys. Rev. B}\
  }\textbf {\bibinfo {volume} {88}},\ \bibinfo {pages} {125438} (\bibinfo
  {year} {2013})}\BibitemShut {NoStop}%
\bibitem [{\citenamefont {Chu}\ \emph {et~al.}(2014)\citenamefont {Chu},
  \citenamefont {Li}, \citenamefont {Wu}, \citenamefont {Niu}, \citenamefont
  {Yao}, \citenamefont {Xu},\ and\ \citenamefont {Zhang}}]{chu2014valley}%
  \BibitemOpen
  \bibfield  {author} {\bibinfo {author} {\bibfnamefont {R.-L.}\ \bibnamefont
  {Chu}}, \bibinfo {author} {\bibfnamefont {X.}~\bibnamefont {Li}}, \bibinfo
  {author} {\bibfnamefont {S.}~\bibnamefont {Wu}}, \bibinfo {author}
  {\bibfnamefont {Q.}~\bibnamefont {Niu}}, \bibinfo {author} {\bibfnamefont
  {W.}~\bibnamefont {Yao}}, \bibinfo {author} {\bibfnamefont {X.}~\bibnamefont
  {Xu}}, \ and\ \bibinfo {author} {\bibfnamefont {C.}~\bibnamefont {Zhang}},\
  }\href@noop {} {\bibfield  {journal} {\bibinfo  {journal} {Phys. Rev. B}\
  }\textbf {\bibinfo {volume} {90}},\ \bibinfo {pages} {045427} (\bibinfo
  {year} {2014})}\BibitemShut {NoStop}%
\bibitem [{\citenamefont {Wang}, \citenamefont {Shan},\ and\ \citenamefont
  {Mak}(2017)}]{wang2017valley}%
  \BibitemOpen
  \bibfield  {author} {\bibinfo {author} {\bibfnamefont {Z.}~\bibnamefont
  {Wang}}, \bibinfo {author} {\bibfnamefont {J.}~\bibnamefont {Shan}}, \ and\
  \bibinfo {author} {\bibfnamefont {K.~F.}\ \bibnamefont {Mak}},\ }\href@noop
  {} {\bibfield  {journal} {\bibinfo  {journal} {Nat. Nanotechnol.}\ }\textbf
  {\bibinfo {volume} {12}},\ \bibinfo {pages} {144} (\bibinfo {year}
  {2017})}\BibitemShut {NoStop}%
\bibitem [{\citenamefont {Srivastava}\ \emph {et~al.}(2015)\citenamefont
  {Srivastava}, \citenamefont {Sidler}, \citenamefont {Allain}, \citenamefont
  {Lembke}, \citenamefont {Kis},\ and\ \citenamefont
  {Imamo{\u{g}}lu}}]{srivastava2015valley}%
  \BibitemOpen
  \bibfield  {author} {\bibinfo {author} {\bibfnamefont {A.}~\bibnamefont
  {Srivastava}}, \bibinfo {author} {\bibfnamefont {M.}~\bibnamefont {Sidler}},
  \bibinfo {author} {\bibfnamefont {A.~V.}\ \bibnamefont {Allain}}, \bibinfo
  {author} {\bibfnamefont {D.~S.}\ \bibnamefont {Lembke}}, \bibinfo {author}
  {\bibfnamefont {A.}~\bibnamefont {Kis}}, \ and\ \bibinfo {author}
  {\bibfnamefont {A.}~\bibnamefont {Imamo{\u{g}}lu}},\ }\href@noop {}
  {\bibfield  {journal} {\bibinfo  {journal} {Nat. Phys.}\ }\textbf {\bibinfo
  {volume} {11}},\ \bibinfo {pages} {141} (\bibinfo {year} {2015})}\BibitemShut
  {NoStop}%
\bibitem [{\citenamefont {Aivazian}\ \emph {et~al.}(2015)\citenamefont
  {Aivazian}, \citenamefont {Gong}, \citenamefont {Jones}, \citenamefont {Chu},
  \citenamefont {Yan}, \citenamefont {Mandrus}, \citenamefont {Zhang},
  \citenamefont {Cobden}, \citenamefont {Yao},\ and\ \citenamefont
  {Xu}}]{aivazian2015magnetic}%
  \BibitemOpen
  \bibfield  {author} {\bibinfo {author} {\bibfnamefont {G.}~\bibnamefont
  {Aivazian}}, \bibinfo {author} {\bibfnamefont {Z.}~\bibnamefont {Gong}},
  \bibinfo {author} {\bibfnamefont {A.~M.}\ \bibnamefont {Jones}}, \bibinfo
  {author} {\bibfnamefont {R.-L.}\ \bibnamefont {Chu}}, \bibinfo {author}
  {\bibfnamefont {J.}~\bibnamefont {Yan}}, \bibinfo {author} {\bibfnamefont
  {D.~G.}\ \bibnamefont {Mandrus}}, \bibinfo {author} {\bibfnamefont
  {C.}~\bibnamefont {Zhang}}, \bibinfo {author} {\bibfnamefont
  {D.}~\bibnamefont {Cobden}}, \bibinfo {author} {\bibfnamefont
  {W.}~\bibnamefont {Yao}}, \ and\ \bibinfo {author} {\bibfnamefont
  {X.}~\bibnamefont {Xu}},\ }\href@noop {} {\bibfield  {journal} {\bibinfo
  {journal} {Nat. Phys.}\ }\textbf {\bibinfo {volume} {11}},\ \bibinfo {pages}
  {148} (\bibinfo {year} {2015})}\BibitemShut {NoStop}%
\bibitem [{\citenamefont {MacNeill}\ \emph {et~al.}(2015)\citenamefont
  {MacNeill}, \citenamefont {Heikes}, \citenamefont {Mak}, \citenamefont
  {Anderson}, \citenamefont {Korm{\'a}nyos}, \citenamefont {Z{\'o}lyomi},
  \citenamefont {Park},\ and\ \citenamefont {Ralph}}]{macneill2015breaking}%
  \BibitemOpen
  \bibfield  {author} {\bibinfo {author} {\bibfnamefont {D.}~\bibnamefont
  {MacNeill}}, \bibinfo {author} {\bibfnamefont {C.}~\bibnamefont {Heikes}},
  \bibinfo {author} {\bibfnamefont {K.~F.}\ \bibnamefont {Mak}}, \bibinfo
  {author} {\bibfnamefont {Z.}~\bibnamefont {Anderson}}, \bibinfo {author}
  {\bibfnamefont {A.}~\bibnamefont {Korm{\'a}nyos}}, \bibinfo {author}
  {\bibfnamefont {V.}~\bibnamefont {Z{\'o}lyomi}}, \bibinfo {author}
  {\bibfnamefont {J.}~\bibnamefont {Park}}, \ and\ \bibinfo {author}
  {\bibfnamefont {D.~C.}\ \bibnamefont {Ralph}},\ }\href@noop {} {\bibfield
  {journal} {\bibinfo  {journal} {Phys. Rev. Lett.}\ }\textbf {\bibinfo
  {volume} {114}},\ \bibinfo {pages} {037401} (\bibinfo {year}
  {2015})}\BibitemShut {NoStop}%
\bibitem [{\citenamefont {Schaibley}\ \emph {et~al.}(2016)\citenamefont
  {Schaibley}, \citenamefont {Yu}, \citenamefont {Clark}, \citenamefont
  {Rivera}, \citenamefont {Ross}, \citenamefont {Seyler}, \citenamefont {Yao},\
  and\ \citenamefont {Xu}}]{schaibley2016valleytronics}%
  \BibitemOpen
  \bibfield  {author} {\bibinfo {author} {\bibfnamefont {J.~R.}\ \bibnamefont
  {Schaibley}}, \bibinfo {author} {\bibfnamefont {H.}~\bibnamefont {Yu}},
  \bibinfo {author} {\bibfnamefont {G.}~\bibnamefont {Clark}}, \bibinfo
  {author} {\bibfnamefont {P.}~\bibnamefont {Rivera}}, \bibinfo {author}
  {\bibfnamefont {J.~S.}\ \bibnamefont {Ross}}, \bibinfo {author}
  {\bibfnamefont {K.~L.}\ \bibnamefont {Seyler}}, \bibinfo {author}
  {\bibfnamefont {W.}~\bibnamefont {Yao}}, \ and\ \bibinfo {author}
  {\bibfnamefont {X.}~\bibnamefont {Xu}},\ }\href@noop {} {\bibfield  {journal}
  {\bibinfo  {journal} {Nat. Rev. Mater.}\ }\textbf {\bibinfo {volume} {1}},\
  \bibinfo {pages} {16055} (\bibinfo {year} {2016})}\BibitemShut {NoStop}%
\bibitem [{\citenamefont {Schmidt}\ \emph {et~al.}(2011)\citenamefont
  {Schmidt}, \citenamefont {Wondraczek}, \citenamefont {Lee}, \citenamefont
  {Granzow}, \citenamefont {Da},\ and\ \citenamefont
  {Russell}}]{schmidt2011complex}%
  \BibitemOpen
  \bibfield  {author} {\bibinfo {author} {\bibfnamefont {M.~A.}\ \bibnamefont
  {Schmidt}}, \bibinfo {author} {\bibfnamefont {L.}~\bibnamefont {Wondraczek}},
  \bibinfo {author} {\bibfnamefont {H.~W.}\ \bibnamefont {Lee}}, \bibinfo
  {author} {\bibfnamefont {N.}~\bibnamefont {Granzow}}, \bibinfo {author}
  {\bibfnamefont {N.}~\bibnamefont {Da}}, \ and\ \bibinfo {author}
  {\bibfnamefont {P.~S.~J.}\ \bibnamefont {Russell}},\ }\href@noop {}
  {\bibfield  {journal} {\bibinfo  {journal} {Adv. Mater.}\ }\textbf {\bibinfo
  {volume} {23}},\ \bibinfo {pages} {2681} (\bibinfo {year}
  {2011})}\BibitemShut {NoStop}%
\bibitem [{\citenamefont {Schmidt}\ \emph {et~al.}(2016)\citenamefont
  {Schmidt}, \citenamefont {Arora}, \citenamefont {Plechinger}, \citenamefont
  {Nagler}, \citenamefont {del {\'A}guila}, \citenamefont {Ballottin},
  \citenamefont {Christianen}, \citenamefont {de~Vasconcellos}, \citenamefont
  {Sch{\"u}ller}, \citenamefont {Korn} \emph {et~al.}}]{schmidt2016magnetic}%
  \BibitemOpen
  \bibfield  {author} {\bibinfo {author} {\bibfnamefont {R.}~\bibnamefont
  {Schmidt}}, \bibinfo {author} {\bibfnamefont {A.}~\bibnamefont {Arora}},
  \bibinfo {author} {\bibfnamefont {G.}~\bibnamefont {Plechinger}}, \bibinfo
  {author} {\bibfnamefont {P.}~\bibnamefont {Nagler}}, \bibinfo {author}
  {\bibfnamefont {A.~G.}\ \bibnamefont {del {\'A}guila}}, \bibinfo {author}
  {\bibfnamefont {M.~V.}\ \bibnamefont {Ballottin}}, \bibinfo {author}
  {\bibfnamefont {P.~C.}\ \bibnamefont {Christianen}}, \bibinfo {author}
  {\bibfnamefont {S.~M.}\ \bibnamefont {de~Vasconcellos}}, \bibinfo {author}
  {\bibfnamefont {C.}~\bibnamefont {Sch{\"u}ller}}, \bibinfo {author}
  {\bibfnamefont {T.}~\bibnamefont {Korn}},  \emph {et~al.},\ }\href@noop {}
  {\bibfield  {journal} {\bibinfo  {journal} {Phys. Rev. Lett.}\ }\textbf
  {\bibinfo {volume} {117}},\ \bibinfo {pages} {077402} (\bibinfo {year}
  {2016})}\BibitemShut {NoStop}%
\bibitem [{\citenamefont {Da}\ \emph {et~al.}(2018)\citenamefont {Da},
  \citenamefont {Gao}, \citenamefont {An}, \citenamefont {Zhang},\ and\
  \citenamefont {Yan}}]{da2018cavity}%
  \BibitemOpen
  \bibfield  {author} {\bibinfo {author} {\bibfnamefont {H.}~\bibnamefont
  {Da}}, \bibinfo {author} {\bibfnamefont {L.}~\bibnamefont {Gao}}, \bibinfo
  {author} {\bibfnamefont {Y.}~\bibnamefont {An}}, \bibinfo {author}
  {\bibfnamefont {H.}~\bibnamefont {Zhang}}, \ and\ \bibinfo {author}
  {\bibfnamefont {X.}~\bibnamefont {Yan}},\ }\href@noop {} {\bibfield
  {journal} {\bibinfo  {journal} {Adv. Opt. Mater.}\ }\textbf {\bibinfo
  {volume} {6}},\ \bibinfo {pages} {1701175} (\bibinfo {year}
  {2018})}\BibitemShut {NoStop}%
\bibitem [{\citenamefont {Stier}\ \emph {et~al.}(2018)\citenamefont {Stier},
  \citenamefont {Wilson}, \citenamefont {Velizhanin}, \citenamefont {Kono},
  \citenamefont {Xu},\ and\ \citenamefont {Crooker}}]{stier2018magnetooptics}%
  \BibitemOpen
  \bibfield  {author} {\bibinfo {author} {\bibfnamefont {A.~V.}\ \bibnamefont
  {Stier}}, \bibinfo {author} {\bibfnamefont {N.~P.}\ \bibnamefont {Wilson}},
  \bibinfo {author} {\bibfnamefont {K.~A.}\ \bibnamefont {Velizhanin}},
  \bibinfo {author} {\bibfnamefont {J.}~\bibnamefont {Kono}}, \bibinfo {author}
  {\bibfnamefont {X.}~\bibnamefont {Xu}}, \ and\ \bibinfo {author}
  {\bibfnamefont {S.~A.}\ \bibnamefont {Crooker}},\ }\href@noop {} {\bibfield
  {journal} {\bibinfo  {journal} {Phys. Rev. Lett.}\ }\textbf {\bibinfo
  {volume} {120}},\ \bibinfo {pages} {057405} (\bibinfo {year}
  {2018})}\BibitemShut {NoStop}%
\bibitem [{\citenamefont {Zipfel}\ \emph {et~al.}(2018)\citenamefont {Zipfel},
  \citenamefont {Holler}, \citenamefont {Mitioglu}, \citenamefont {Ballottin},
  \citenamefont {Nagler}, \citenamefont {Stier}, \citenamefont {Taniguchi},
  \citenamefont {Watanabe}, \citenamefont {Crooker}, \citenamefont
  {Christianen} \emph {et~al.}}]{zipfel2018spatial}%
  \BibitemOpen
  \bibfield  {author} {\bibinfo {author} {\bibfnamefont {J.}~\bibnamefont
  {Zipfel}}, \bibinfo {author} {\bibfnamefont {J.}~\bibnamefont {Holler}},
  \bibinfo {author} {\bibfnamefont {A.~A.}\ \bibnamefont {Mitioglu}}, \bibinfo
  {author} {\bibfnamefont {M.~V.}\ \bibnamefont {Ballottin}}, \bibinfo {author}
  {\bibfnamefont {P.}~\bibnamefont {Nagler}}, \bibinfo {author} {\bibfnamefont
  {A.~V.}\ \bibnamefont {Stier}}, \bibinfo {author} {\bibfnamefont
  {T.}~\bibnamefont {Taniguchi}}, \bibinfo {author} {\bibfnamefont
  {K.}~\bibnamefont {Watanabe}}, \bibinfo {author} {\bibfnamefont {S.~A.}\
  \bibnamefont {Crooker}}, \bibinfo {author} {\bibfnamefont {P.~C.}\
  \bibnamefont {Christianen}},  \emph {et~al.},\ }\href@noop {} {\bibfield
  {journal} {\bibinfo  {journal} {Phys. Rev. B}\ }\textbf {\bibinfo {volume}
  {98}},\ \bibinfo {pages} {075438} (\bibinfo {year} {2018})}\BibitemShut
  {NoStop}%
\bibitem [{\citenamefont {Stier}\ \emph
  {et~al.}(2016{\natexlab{a}})\citenamefont {Stier}, \citenamefont {Wilson},
  \citenamefont {Clark}, \citenamefont {Xu},\ and\ \citenamefont
  {Crooker}}]{stier2016probing}%
  \BibitemOpen
  \bibfield  {author} {\bibinfo {author} {\bibfnamefont {A.~V.}\ \bibnamefont
  {Stier}}, \bibinfo {author} {\bibfnamefont {N.~P.}\ \bibnamefont {Wilson}},
  \bibinfo {author} {\bibfnamefont {G.}~\bibnamefont {Clark}}, \bibinfo
  {author} {\bibfnamefont {X.}~\bibnamefont {Xu}}, \ and\ \bibinfo {author}
  {\bibfnamefont {S.~A.}\ \bibnamefont {Crooker}},\ }\href@noop {} {\bibfield
  {journal} {\bibinfo  {journal} {Nano Lett.}\ }\textbf {\bibinfo {volume}
  {16}},\ \bibinfo {pages} {7054} (\bibinfo {year}
  {2016}{\natexlab{a}})}\BibitemShut {NoStop}%
\bibitem [{\citenamefont {Mitioglu}\ \emph {et~al.}(2016)\citenamefont
  {Mitioglu}, \citenamefont {Galkowski}, \citenamefont {Surrente},
  \citenamefont {Klopotowski}, \citenamefont {Dumcenco}, \citenamefont {Kis},
  \citenamefont {Maude},\ and\ \citenamefont
  {Plochocka}}]{mitioglu2016magnetoexcitons}%
  \BibitemOpen
  \bibfield  {author} {\bibinfo {author} {\bibfnamefont {A.}~\bibnamefont
  {Mitioglu}}, \bibinfo {author} {\bibfnamefont {K.}~\bibnamefont {Galkowski}},
  \bibinfo {author} {\bibfnamefont {A.}~\bibnamefont {Surrente}}, \bibinfo
  {author} {\bibfnamefont {L.}~\bibnamefont {Klopotowski}}, \bibinfo {author}
  {\bibfnamefont {D.}~\bibnamefont {Dumcenco}}, \bibinfo {author}
  {\bibfnamefont {A.}~\bibnamefont {Kis}}, \bibinfo {author} {\bibfnamefont
  {D.}~\bibnamefont {Maude}}, \ and\ \bibinfo {author} {\bibfnamefont
  {P.}~\bibnamefont {Plochocka}},\ }\href@noop {} {\bibfield  {journal}
  {\bibinfo  {journal} {Phys. Rev. B}\ }\textbf {\bibinfo {volume} {93}},\
  \bibinfo {pages} {165412} (\bibinfo {year} {2016})}\BibitemShut {NoStop}%
\bibitem [{\citenamefont {Stier}\ \emph
  {et~al.}(2016{\natexlab{b}})\citenamefont {Stier}, \citenamefont {McCreary},
  \citenamefont {Jonker}, \citenamefont {Kono},\ and\ \citenamefont
  {Crooker}}]{stier2016exciton}%
  \BibitemOpen
  \bibfield  {author} {\bibinfo {author} {\bibfnamefont {A.~V.}\ \bibnamefont
  {Stier}}, \bibinfo {author} {\bibfnamefont {K.~M.}\ \bibnamefont {McCreary}},
  \bibinfo {author} {\bibfnamefont {B.~T.}\ \bibnamefont {Jonker}}, \bibinfo
  {author} {\bibfnamefont {J.}~\bibnamefont {Kono}}, \ and\ \bibinfo {author}
  {\bibfnamefont {S.~A.}\ \bibnamefont {Crooker}},\ }\href@noop {} {\bibfield
  {journal} {\bibinfo  {journal} {Nat. Commun.}\ }\textbf {\bibinfo {volume}
  {7}},\ \bibinfo {pages} {10643} (\bibinfo {year}
  {2016}{\natexlab{b}})}\BibitemShut {NoStop}%
\bibitem [{\citenamefont {Li}\ \emph {et~al.}(2014)\citenamefont {Li},
  \citenamefont {Ludwig}, \citenamefont {Low}, \citenamefont {Chernikov},
  \citenamefont {Cui}, \citenamefont {Arefe}, \citenamefont {Kim},
  \citenamefont {van~der Zande}, \citenamefont {Rigosi}, \citenamefont {Hill}
  \emph {et~al.}}]{li2014valley}%
  \BibitemOpen
  \bibfield  {author} {\bibinfo {author} {\bibfnamefont {Y.}~\bibnamefont
  {Li}}, \bibinfo {author} {\bibfnamefont {J.}~\bibnamefont {Ludwig}}, \bibinfo
  {author} {\bibfnamefont {T.}~\bibnamefont {Low}}, \bibinfo {author}
  {\bibfnamefont {A.}~\bibnamefont {Chernikov}}, \bibinfo {author}
  {\bibfnamefont {X.}~\bibnamefont {Cui}}, \bibinfo {author} {\bibfnamefont
  {G.}~\bibnamefont {Arefe}}, \bibinfo {author} {\bibfnamefont {Y.~D.}\
  \bibnamefont {Kim}}, \bibinfo {author} {\bibfnamefont {A.~M.}\ \bibnamefont
  {van~der Zande}}, \bibinfo {author} {\bibfnamefont {A.}~\bibnamefont
  {Rigosi}}, \bibinfo {author} {\bibfnamefont {H.~M.}\ \bibnamefont {Hill}},
  \emph {et~al.},\ }\href@noop {} {\bibfield  {journal} {\bibinfo  {journal}
  {Phys. Rev. Lett.}\ }\textbf {\bibinfo {volume} {113}},\ \bibinfo {pages}
  {266804} (\bibinfo {year} {2014})}\BibitemShut {NoStop}%
\bibitem [{\citenamefont {Plechinger}\ \emph {et~al.}(2016)\citenamefont
  {Plechinger}, \citenamefont {Nagler}, \citenamefont {Arora}, \citenamefont
  {Granados~del Aguila}, \citenamefont {Ballottin}, \citenamefont {Frank},
  \citenamefont {Steinleitner}, \citenamefont {Gmitra}, \citenamefont {Fabian},
  \citenamefont {Christianen} \emph {et~al.}}]{plechinger2016excitonic}%
  \BibitemOpen
  \bibfield  {author} {\bibinfo {author} {\bibfnamefont {G.}~\bibnamefont
  {Plechinger}}, \bibinfo {author} {\bibfnamefont {P.}~\bibnamefont {Nagler}},
  \bibinfo {author} {\bibfnamefont {A.}~\bibnamefont {Arora}}, \bibinfo
  {author} {\bibfnamefont {A.}~\bibnamefont {Granados~del Aguila}}, \bibinfo
  {author} {\bibfnamefont {M.~V.}\ \bibnamefont {Ballottin}}, \bibinfo {author}
  {\bibfnamefont {T.}~\bibnamefont {Frank}}, \bibinfo {author} {\bibfnamefont
  {P.}~\bibnamefont {Steinleitner}}, \bibinfo {author} {\bibfnamefont
  {M.}~\bibnamefont {Gmitra}}, \bibinfo {author} {\bibfnamefont
  {J.}~\bibnamefont {Fabian}}, \bibinfo {author} {\bibfnamefont {P.~C.}\
  \bibnamefont {Christianen}},  \emph {et~al.},\ }\href@noop {} {\bibfield
  {journal} {\bibinfo  {journal} {Nano Lett.}\ }\textbf {\bibinfo {volume}
  {16}},\ \bibinfo {pages} {7899} (\bibinfo {year} {2016})}\BibitemShut
  {NoStop}%
\bibitem [{\citenamefont {Stier}\ \emph
  {et~al.}(2016{\natexlab{c}})\citenamefont {Stier}, \citenamefont {McCreary},
  \citenamefont {Jonker}, \citenamefont {Kono},\ and\ \citenamefont
  {Crooker}}]{stier2016magnetoreflection}%
  \BibitemOpen
  \bibfield  {author} {\bibinfo {author} {\bibfnamefont {A.~V.}\ \bibnamefont
  {Stier}}, \bibinfo {author} {\bibfnamefont {K.~M.}\ \bibnamefont {McCreary}},
  \bibinfo {author} {\bibfnamefont {B.~T.}\ \bibnamefont {Jonker}}, \bibinfo
  {author} {\bibfnamefont {J.}~\bibnamefont {Kono}}, \ and\ \bibinfo {author}
  {\bibfnamefont {S.~A.}\ \bibnamefont {Crooker}},\ }\href@noop {} {\bibfield
  {journal} {\bibinfo  {journal} {J. Vac. Sci. Technol.}\ }\textbf {\bibinfo
  {volume} {34}},\ \bibinfo {pages} {04J102} (\bibinfo {year}
  {2016}{\natexlab{c}})}\BibitemShut {NoStop}%
\bibitem [{\citenamefont {Mitioglu}\ \emph {et~al.}(2015)\citenamefont
  {Mitioglu}, \citenamefont {Plochocka}, \citenamefont {Granados~del Aguila},
  \citenamefont {Christianen}, \citenamefont {Deligeorgis}, \citenamefont
  {Anghel}, \citenamefont {Kulyuk},\ and\ \citenamefont
  {Maude}}]{mitioglu2015optical}%
  \BibitemOpen
  \bibfield  {author} {\bibinfo {author} {\bibfnamefont {A.}~\bibnamefont
  {Mitioglu}}, \bibinfo {author} {\bibfnamefont {P.}~\bibnamefont {Plochocka}},
  \bibinfo {author} {\bibfnamefont {A.}~\bibnamefont {Granados~del Aguila}},
  \bibinfo {author} {\bibfnamefont {P.}~\bibnamefont {Christianen}}, \bibinfo
  {author} {\bibfnamefont {G.}~\bibnamefont {Deligeorgis}}, \bibinfo {author}
  {\bibfnamefont {S.}~\bibnamefont {Anghel}}, \bibinfo {author} {\bibfnamefont
  {L.}~\bibnamefont {Kulyuk}}, \ and\ \bibinfo {author} {\bibfnamefont
  {D.}~\bibnamefont {Maude}},\ }\href@noop {} {\bibfield  {journal} {\bibinfo
  {journal} {Nano Lett.}\ }\textbf {\bibinfo {volume} {15}},\ \bibinfo {pages}
  {4387} (\bibinfo {year} {2015})}\BibitemShut {NoStop}%
\bibitem [{\citenamefont {Have}\ and\ \citenamefont
  {Pedersen}(2018)}]{have2018magnetoexcitons}%
  \BibitemOpen
  \bibfield  {author} {\bibinfo {author} {\bibfnamefont {J.}~\bibnamefont
  {Have}}\ and\ \bibinfo {author} {\bibfnamefont {T.~G.}\ \bibnamefont
  {Pedersen}},\ }\href@noop {} {\bibfield  {journal} {\bibinfo  {journal}
  {Phys. Rev. B}\ }\textbf {\bibinfo {volume} {97}},\ \bibinfo {pages} {115405}
  (\bibinfo {year} {2018})}\BibitemShut {NoStop}%
\bibitem [{\citenamefont {Wannier}(1937)}]{wannier1937structure}%
  \BibitemOpen
  \bibfield  {author} {\bibinfo {author} {\bibfnamefont {G.~H.}\ \bibnamefont
  {Wannier}},\ }\href@noop {} {\bibfield  {journal} {\bibinfo  {journal} {Phys.
  Rev.}\ }\textbf {\bibinfo {volume} {52}},\ \bibinfo {pages} {191} (\bibinfo
  {year} {1937})}\BibitemShut {NoStop}%
\bibitem [{\citenamefont {Van~der Donck}, \citenamefont {Zarenia},\ and\
  \citenamefont {Peeters}(2018)}]{van2018excitons}%
  \BibitemOpen
  \bibfield  {author} {\bibinfo {author} {\bibfnamefont {M.}~\bibnamefont
  {Van~der Donck}}, \bibinfo {author} {\bibfnamefont {M.}~\bibnamefont
  {Zarenia}}, \ and\ \bibinfo {author} {\bibfnamefont {F.}~\bibnamefont
  {Peeters}},\ }\href@noop {} {\bibfield  {journal} {\bibinfo  {journal} {Phys.
  Rev. B}\ }\textbf {\bibinfo {volume} {97}},\ \bibinfo {pages} {195408}
  (\bibinfo {year} {2018})}\BibitemShut {NoStop}%
\bibitem [{\citenamefont {Pedersen}(2015)}]{pedersen2015intraband}%
  \BibitemOpen
  \bibfield  {author} {\bibinfo {author} {\bibfnamefont {T.~G.}\ \bibnamefont
  {Pedersen}},\ }\href@noop {} {\bibfield  {journal} {\bibinfo  {journal}
  {Phys. Rev. B}\ }\textbf {\bibinfo {volume} {92}},\ \bibinfo {pages} {235432}
  (\bibinfo {year} {2015})}\BibitemShut {NoStop}%
\bibitem [{\citenamefont {Liu}\ \emph {et~al.}(2013)\citenamefont {Liu},
  \citenamefont {Shan}, \citenamefont {Yao}, \citenamefont {Yao},\ and\
  \citenamefont {Xiao}}]{liu2013three}%
  \BibitemOpen
  \bibfield  {author} {\bibinfo {author} {\bibfnamefont {G.-B.}\ \bibnamefont
  {Liu}}, \bibinfo {author} {\bibfnamefont {W.-Y.}\ \bibnamefont {Shan}},
  \bibinfo {author} {\bibfnamefont {Y.}~\bibnamefont {Yao}}, \bibinfo {author}
  {\bibfnamefont {W.}~\bibnamefont {Yao}}, \ and\ \bibinfo {author}
  {\bibfnamefont {D.}~\bibnamefont {Xiao}},\ }\href@noop {} {\bibfield
  {journal} {\bibinfo  {journal} {Phys. Rev. B}\ }\textbf {\bibinfo {volume}
  {88}},\ \bibinfo {pages} {085433} (\bibinfo {year} {2013})}\BibitemShut
  {NoStop}%
\bibitem [{\citenamefont {Cheng}\ and\ \citenamefont
  {Guo}(2018)}]{cheng2018nonlinear}%
  \BibitemOpen
  \bibfield  {author} {\bibinfo {author} {\bibfnamefont {J.}~\bibnamefont
  {Cheng}}\ and\ \bibinfo {author} {\bibfnamefont {C.}~\bibnamefont {Guo}},\
  }\href@noop {} {\bibfield  {journal} {\bibinfo  {journal} {Phys. Rev. B}\
  }\textbf {\bibinfo {volume} {97}},\ \bibinfo {pages} {125417} (\bibinfo
  {year} {2018})}\BibitemShut {NoStop}%
\bibitem [{\citenamefont {Rybkovskiy}, \citenamefont {Gerber},\ and\
  \citenamefont {Durnev}(2017)}]{rybkovskiy2017atomically}%
  \BibitemOpen
  \bibfield  {author} {\bibinfo {author} {\bibfnamefont {D.}~\bibnamefont
  {Rybkovskiy}}, \bibinfo {author} {\bibfnamefont {I.}~\bibnamefont {Gerber}},
  \ and\ \bibinfo {author} {\bibfnamefont {M.}~\bibnamefont {Durnev}},\
  }\href@noop {} {\bibfield  {journal} {\bibinfo  {journal} {Phys. Rev. B}\
  }\textbf {\bibinfo {volume} {95}},\ \bibinfo {pages} {155406} (\bibinfo
  {year} {2017})}\BibitemShut {NoStop}%
\bibitem [{\citenamefont {Ferreira}\ \emph {et~al.}(2011)\citenamefont
  {Ferreira}, \citenamefont {Viana-Gomes}, \citenamefont {Bludov},
  \citenamefont {Pereira}, \citenamefont {Peres},\ and\ \citenamefont
  {Neto}}]{ferreira2011faraday}%
  \BibitemOpen
  \bibfield  {author} {\bibinfo {author} {\bibfnamefont {A.}~\bibnamefont
  {Ferreira}}, \bibinfo {author} {\bibfnamefont {J.}~\bibnamefont
  {Viana-Gomes}}, \bibinfo {author} {\bibfnamefont {Y.~V.}\ \bibnamefont
  {Bludov}}, \bibinfo {author} {\bibfnamefont {V.}~\bibnamefont {Pereira}},
  \bibinfo {author} {\bibfnamefont {N.}~\bibnamefont {Peres}}, \ and\ \bibinfo
  {author} {\bibfnamefont {A.~C.}\ \bibnamefont {Neto}},\ }\href@noop {}
  {\bibfield  {journal} {\bibinfo  {journal} {Phys. Rev. B}\ }\textbf {\bibinfo
  {volume} {84}},\ \bibinfo {pages} {235410} (\bibinfo {year}
  {2011})}\BibitemShut {NoStop}%
\bibitem [{\citenamefont {Catarina}\ \emph {et~al.}(2018)\citenamefont
  {Catarina}, \citenamefont {Have}, \citenamefont {Fernandez-Rossier},\ and\
  \citenamefont {Peres}}]{catarina2018magneto}%
  \BibitemOpen
  \bibfield  {author} {\bibinfo {author} {\bibfnamefont {G.}~\bibnamefont
  {Catarina}}, \bibinfo {author} {\bibfnamefont {J.}~\bibnamefont {Have}},
  \bibinfo {author} {\bibfnamefont {J.}~\bibnamefont {Fernandez-Rossier}}, \
  and\ \bibinfo {author} {\bibfnamefont {N.~M.~R.}\ \bibnamefont {Peres}},\
  }\href@noop {} {\bibfield  {journal} {\bibinfo  {journal} {In preperation}\ }
  (\bibinfo {year} {2018})}\BibitemShut {NoStop}%
\bibitem [{\citenamefont {Keldysh}(1979)}]{keldysh1979coulomb}%
  \BibitemOpen
  \bibfield  {author} {\bibinfo {author} {\bibfnamefont {L.}~\bibnamefont
  {Keldysh}},\ }\href@noop {} {\bibfield  {journal} {\bibinfo  {journal} {Sov.
  Phys. JETP}\ }\textbf {\bibinfo {volume} {29}},\ \bibinfo {pages} {658}
  (\bibinfo {year} {1979})}\BibitemShut {NoStop}%
\bibitem [{\citenamefont {Cudazzo}, \citenamefont {Tokatly},\ and\
  \citenamefont {Rubio}(2011)}]{cudazzo2011dielectric}%
  \BibitemOpen
  \bibfield  {author} {\bibinfo {author} {\bibfnamefont {P.}~\bibnamefont
  {Cudazzo}}, \bibinfo {author} {\bibfnamefont {I.~V.}\ \bibnamefont
  {Tokatly}}, \ and\ \bibinfo {author} {\bibfnamefont {A.}~\bibnamefont
  {Rubio}},\ }\href@noop {} {\bibfield  {journal} {\bibinfo  {journal} {Phys.
  Rev. B}\ }\textbf {\bibinfo {volume} {84}},\ \bibinfo {pages} {085406}
  (\bibinfo {year} {2011})}\BibitemShut {NoStop}%
\bibitem [{\citenamefont {Trolle}, \citenamefont {Pedersen},\ and\
  \citenamefont {V{\'e}niard}(2017)}]{trolle2017model}%
  \BibitemOpen
  \bibfield  {author} {\bibinfo {author} {\bibfnamefont {M.~L.}\ \bibnamefont
  {Trolle}}, \bibinfo {author} {\bibfnamefont {T.~G.}\ \bibnamefont
  {Pedersen}}, \ and\ \bibinfo {author} {\bibfnamefont {V.}~\bibnamefont
  {V{\'e}niard}},\ }\href@noop {} {\bibfield  {journal} {\bibinfo  {journal}
  {Sci. Rep.}\ }\textbf {\bibinfo {volume} {7}},\ \bibinfo {pages} {39844}
  (\bibinfo {year} {2017})}\BibitemShut {NoStop}%
\bibitem [{\citenamefont {Ehrenreich}\ and\ \citenamefont
  {Cohen}(1959)}]{ehrenreich1959self}%
  \BibitemOpen
  \bibfield  {author} {\bibinfo {author} {\bibfnamefont {H.}~\bibnamefont
  {Ehrenreich}}\ and\ \bibinfo {author} {\bibfnamefont {M.~H.}\ \bibnamefont
  {Cohen}},\ }\href@noop {} {\bibfield  {journal} {\bibinfo  {journal} {Phys.
  Rev.}\ }\textbf {\bibinfo {volume} {115}},\ \bibinfo {pages} {786} (\bibinfo
  {year} {1959})}\BibitemShut {NoStop}%
\bibitem [{\citenamefont {Shizuya}(2010)}]{shizuya2010many}%
  \BibitemOpen
  \bibfield  {author} {\bibinfo {author} {\bibfnamefont {K.}~\bibnamefont
  {Shizuya}},\ }\href@noop {} {\bibfield  {journal} {\bibinfo  {journal} {Phys.
  Rev. B}\ }\textbf {\bibinfo {volume} {81}},\ \bibinfo {pages} {075407}
  (\bibinfo {year} {2010})}\BibitemShut {NoStop}%
\bibitem [{\citenamefont {Sokolik}, \citenamefont {Zabolotskiy},\ and\
  \citenamefont {Lozovik}(2017)}]{sokolik2017many}%
  \BibitemOpen
  \bibfield  {author} {\bibinfo {author} {\bibfnamefont {A.}~\bibnamefont
  {Sokolik}}, \bibinfo {author} {\bibfnamefont {A.}~\bibnamefont
  {Zabolotskiy}}, \ and\ \bibinfo {author} {\bibfnamefont {Y.~E.}\ \bibnamefont
  {Lozovik}},\ }\href@noop {} {\bibfield  {journal} {\bibinfo  {journal} {Phys.
  Rev. B}\ }\textbf {\bibinfo {volume} {95}},\ \bibinfo {pages} {125402}
  (\bibinfo {year} {2017})}\BibitemShut {NoStop}%
\bibitem [{\citenamefont {Nilsson}\ \emph {et~al.}(2006)\citenamefont
  {Nilsson}, \citenamefont {Neto}, \citenamefont {Guinea},\ and\ \citenamefont
  {Peres}}]{nilsson2006electronic}%
  \BibitemOpen
  \bibfield  {author} {\bibinfo {author} {\bibfnamefont {J.}~\bibnamefont
  {Nilsson}}, \bibinfo {author} {\bibfnamefont {A.~C.}\ \bibnamefont {Neto}},
  \bibinfo {author} {\bibfnamefont {F.}~\bibnamefont {Guinea}}, \ and\ \bibinfo
  {author} {\bibfnamefont {N.}~\bibnamefont {Peres}},\ }\href@noop {}
  {\bibfield  {journal} {\bibinfo  {journal} {Phys. Rev. Lett.}\ }\textbf
  {\bibinfo {volume} {97}},\ \bibinfo {pages} {266801} (\bibinfo {year}
  {2006})}\BibitemShut {NoStop}%
\bibitem [{\citenamefont {Rasmussen}\ and\ \citenamefont
  {Thygesen}(2015)}]{rasmussen2015computational}%
  \BibitemOpen
  \bibfield  {author} {\bibinfo {author} {\bibfnamefont {F.~A.}\ \bibnamefont
  {Rasmussen}}\ and\ \bibinfo {author} {\bibfnamefont {K.~S.}\ \bibnamefont
  {Thygesen}},\ }\href@noop {} {\bibfield  {journal} {\bibinfo  {journal} {J.
  Phys. Chem. C}\ }\textbf {\bibinfo {volume} {119}},\ \bibinfo {pages} {13169}
  (\bibinfo {year} {2015})}\BibitemShut {NoStop}%
\bibitem [{\citenamefont {Rohlfing}\ and\ \citenamefont
  {Louie}(2000)}]{rohlfing2000electron}%
  \BibitemOpen
  \bibfield  {author} {\bibinfo {author} {\bibfnamefont {M.}~\bibnamefont
  {Rohlfing}}\ and\ \bibinfo {author} {\bibfnamefont {S.~G.}\ \bibnamefont
  {Louie}},\ }\href@noop {} {\bibfield  {journal} {\bibinfo  {journal} {Phys.
  Rev. B}\ }\textbf {\bibinfo {volume} {62}},\ \bibinfo {pages} {4927}
  (\bibinfo {year} {2000})}\BibitemShut {NoStop}%
\bibitem [{\citenamefont {Stafford}, \citenamefont {Schmitt-Rink},\ and\
  \citenamefont {Schaefer}(1990)}]{stafford1990nonlinear}%
  \BibitemOpen
  \bibfield  {author} {\bibinfo {author} {\bibfnamefont {C.}~\bibnamefont
  {Stafford}}, \bibinfo {author} {\bibfnamefont {S.}~\bibnamefont
  {Schmitt-Rink}}, \ and\ \bibinfo {author} {\bibfnamefont {W.}~\bibnamefont
  {Schaefer}},\ }\href@noop {} {\bibfield  {journal} {\bibinfo  {journal}
  {Phys. Rev. B}\ }\textbf {\bibinfo {volume} {41}},\ \bibinfo {pages} {10000}
  (\bibinfo {year} {1990})}\BibitemShut {NoStop}%
\bibitem [{\citenamefont {Pedersen}(2016)}]{pedersen2016exciton}%
  \BibitemOpen
  \bibfield  {author} {\bibinfo {author} {\bibfnamefont {T.~G.}\ \bibnamefont
  {Pedersen}},\ }\href@noop {} {\bibfield  {journal} {\bibinfo  {journal}
  {Phys. Rev. B}\ }\textbf {\bibinfo {volume} {94}},\ \bibinfo {pages} {125424}
  (\bibinfo {year} {2016})}\BibitemShut {NoStop}%
\bibitem [{\citenamefont {Massicotte}\ \emph {et~al.}(2018)\citenamefont
  {Massicotte}, \citenamefont {Vialla}, \citenamefont {Schmidt}, \citenamefont
  {Lundeberg}, \citenamefont {Latini}, \citenamefont {Haastrup}, \citenamefont
  {Danovich}, \citenamefont {Davydovskaya}, \citenamefont {Watanabe},
  \citenamefont {Taniguchi} \emph {et~al.}}]{massicotte2018dissociation}%
  \BibitemOpen
  \bibfield  {author} {\bibinfo {author} {\bibfnamefont {M.}~\bibnamefont
  {Massicotte}}, \bibinfo {author} {\bibfnamefont {F.}~\bibnamefont {Vialla}},
  \bibinfo {author} {\bibfnamefont {P.}~\bibnamefont {Schmidt}}, \bibinfo
  {author} {\bibfnamefont {M.~B.}\ \bibnamefont {Lundeberg}}, \bibinfo {author}
  {\bibfnamefont {S.}~\bibnamefont {Latini}}, \bibinfo {author} {\bibfnamefont
  {S.}~\bibnamefont {Haastrup}}, \bibinfo {author} {\bibfnamefont
  {M.}~\bibnamefont {Danovich}}, \bibinfo {author} {\bibfnamefont
  {D.}~\bibnamefont {Davydovskaya}}, \bibinfo {author} {\bibfnamefont
  {K.}~\bibnamefont {Watanabe}}, \bibinfo {author} {\bibfnamefont
  {T.}~\bibnamefont {Taniguchi}},  \emph {et~al.},\ }\href@noop {} {\bibfield
  {journal} {\bibinfo  {journal} {Nature Commun.}\ }\textbf {\bibinfo {volume}
  {9}},\ \bibinfo {pages} {1633} (\bibinfo {year} {2018})}\BibitemShut
  {NoStop}%
\bibitem [{\citenamefont {Johansson}\ \emph {et~al.}(2013)\citenamefont
  {Johansson} \emph {et~al.}}]{mpmath}%
  \BibitemOpen
  \bibfield  {author} {\bibinfo {author} {\bibfnamefont {F.}~\bibnamefont
  {Johansson}} \emph {et~al.},\ }\href@noop {} {\emph {\bibinfo {title}
  {mpmath: a {P}ython library for arbitrary-precision floating-point arithmetic
  (version 0.18)}}} (\bibinfo {year} {2013}),\ \bibinfo {note} {{\tt
  http://mpmath.org/}}\BibitemShut {NoStop}%
\bibitem [{\citenamefont {Lin}\ \emph {et~al.}(2014)\citenamefont {Lin},
  \citenamefont {Ling}, \citenamefont {Yu}, \citenamefont {Huang},
  \citenamefont {Hsu}, \citenamefont {Lee}, \citenamefont {Kong}, \citenamefont
  {Dresselhaus},\ and\ \citenamefont {Palacios}}]{lin2014dielectric}%
  \BibitemOpen
  \bibfield  {author} {\bibinfo {author} {\bibfnamefont {Y.}~\bibnamefont
  {Lin}}, \bibinfo {author} {\bibfnamefont {X.}~\bibnamefont {Ling}}, \bibinfo
  {author} {\bibfnamefont {L.}~\bibnamefont {Yu}}, \bibinfo {author}
  {\bibfnamefont {S.}~\bibnamefont {Huang}}, \bibinfo {author} {\bibfnamefont
  {A.~L.}\ \bibnamefont {Hsu}}, \bibinfo {author} {\bibfnamefont {Y.-H.}\
  \bibnamefont {Lee}}, \bibinfo {author} {\bibfnamefont {J.}~\bibnamefont
  {Kong}}, \bibinfo {author} {\bibfnamefont {M.~S.}\ \bibnamefont
  {Dresselhaus}}, \ and\ \bibinfo {author} {\bibfnamefont {T.}~\bibnamefont
  {Palacios}},\ }\href@noop {} {\bibfield  {journal} {\bibinfo  {journal} {Nano
  Lett.}\ }\textbf {\bibinfo {volume} {14}},\ \bibinfo {pages} {5569} (\bibinfo
  {year} {2014})}\BibitemShut {NoStop}%
\bibitem [{\citenamefont {Mahan}(2013)}]{mahan2013many}%
  \BibitemOpen
  \bibfield  {author} {\bibinfo {author} {\bibfnamefont {G.~D.}\ \bibnamefont
  {Mahan}},\ }\href@noop {} {\emph {\bibinfo {title} {Many-particle physics}}}\
  (\bibinfo  {publisher} {Springer Science \& Business Media},\ \bibinfo {year}
  {2013})\BibitemShut {NoStop}%
\bibitem [{\citenamefont {Bychkov}\ and\ \citenamefont
  {Rashba}(1983)}]{bychkov1983two}%
  \BibitemOpen
  \bibfield  {author} {\bibinfo {author} {\bibfnamefont {I.}~\bibnamefont
  {Bychkov}}\ and\ \bibinfo {author} {\bibfnamefont {E.}~\bibnamefont
  {Rashba}},\ }\href@noop {} {\bibfield  {journal} {\bibinfo  {journal} {Zh.
  Eksp. Teor. Fiz.}\ }\textbf {\bibinfo {volume} {85}},\ \bibinfo {pages}
  {1826} (\bibinfo {year} {1983})}\BibitemShut {NoStop}%
\end{thebibliography}%
\end{document}